\documentclass[a4paper,12pt,authoryear]{elsarticle}
\usepackage{graphicx}
\usepackage{geometry}
\usepackage{amsmath}
\usepackage{amssymb}
\usepackage{natbib}
\usepackage{graphics}
\usepackage{color}
\usepackage[normalsize]{subfigure}
\usepackage{hyperref}
\usepackage{breakcites}

\usepackage{amsmath}
\usepackage{amsfonts}

\newcommand{\DOFs}{DOFs}

\newcommand{\bmath}[1]{{\boldsymbol #1}}

\newcommand{\vek}[1]{\bmath{#1}} 
\newcommand{\mtrx}[1]{\bmath{#1}} 
\newcommand{\trn}{{^{\sf T}}}

\DeclareMathOperator*{\assembly}{{\sf A}}

\DeclareMathOperator*{\stat}{\mathrm{stat}}

\newcommand{\el}{e} 
\newcommand{\fel}{_{\el}} 
\newcommand{\numel}{{N_\el}} 
\newcommand{\numno}{{N_n}} 
\newcommand{\numdof}{{N_d}} 
\newcommand{\numstress
}{{N_s}} 
\newcommand{\kM}{\mtrx{B}} 
\newcommand{\sM}{\mtrx{B}\trn} 
\newcommand{\mM}{\mtrx{D}} 
\newcommand{\dV}{\vek{\dof}} 
\newcommand{\sV}{\vek{s}} 
\newcommand{\eV}{\vek{e}} 
\newcommand{\fV}{\vek{f}} 
\newcommand{\tV}{\vek{\tau}} 
\newcommand{\stM}{\mtrx{K}}

\newcommand{\gM}{\mtrx{\Gamma}}
\newcommand{\cF}{\chi}
\newcommand{\cV}{\vek{\chi}}
\newcommand{\rl}{\alpha} 
\newcommand{\avg}[2]{\left\langle #1 \right\rangle_{#2}}

\newcommand{\enspc}{\set{S}}
 
\newcommand{\oppF}{p}
\newcommand{\tppF}{P}

\newcommand{\UF}{U}

\newcommand{\EF}{E}

\newcommand{\trial}[1]{\widehat{#1}}
\newcommand{\phs}[1]{^{(#1)}} 

\newcommand{\tR}{\kappa} 

\newcommand{\pV}{\vek{p}}
\newcommand{\PM}{\mtrx{P}}

\newcommand{\pd}{\mu}
\newcommand{\pdV}{\vek{\mu}}

\newcommand{\inv}{^{-1}}

\newcommand{\HF}{H}
\newcommand{\sHF}{{\mathcal H}}

\renewcommand{\eqref}[1]{(\ref{eq:#1})}
\newcommand{\Fref}[1]{\figurename~\ref{fig:#1}}

\newcommand{\Eref}[1]{Eq.~\eqref{#1}}
\newcommand{\Sref}[1]{Section~\ref{sec:#1}}

\newcommand{\sprod}{\bullet} 

\newcommand{\myMatrix}[2]{%
\left[
\begin{array}{#1}%
#2%
\end{array}%
\right]
}
 
\newcommand{\admPD}{\set{M}}
\newcommand{\bqp}{\set{B}_Q}

\newcommand{\free}{\vek{u}}
\newcommand{\fixed}{\vek{c}}
\newcommand{\dof}{d}
\newcommand{\set}[1]{{\mathbb #1}}

\newcommand{\card}[1]{|#1|}

\newcommand{\diag}[1]{\mathrm{diag}(#1)}

\newcommand{\half}{\mbox{$\frac{1}{2}$}}

\newcommand{\beq}{\begin{equation}}
\newcommand{\eeq}{\end{equation}}

\newcommand{\cmp}[1]{{_{(#1)}}}

\newcommand{\smplx}{\Delta}
\newcommand{\up}{_{+}}
\newcommand{\low}{_{-}}
\newcommand{\appr}{_{\sim}}

\newcommand{\uM}{\mtrx{I}} 
\newcommand{\sopt}[1]{\tilde{#1}} 

\newcommand{\FE}{\ensuremath{\mathrm{FE}}}

\newcommand{\pma}{\stackrel{\pm}{\sim}}
\newcommand{\lessgrta}{\mbox{$\lessgtr \atop \sim$}}

\newcommand{\FC}{\mathrm{FC}}
\newcommand{\DC}{\mathrm{DC}}
\newcommand{\R}{\mathrm{R}}
\newcommand{\V}{\mathrm{V}}
\newcommand{\MC}{\mathrm{MC}}
\newcommand{\HSW}{\mathrm{HSW}}
\newcommand{\relE}{\zeta} 
\newcommand{\crt}{^{\mathrm{cr}}}
\newcommand{\by}[2]{\stackrel{\eqref{#1}}{#2}}

\journal{Journal of the Mechanics and Physics of Solids}

\begin{document}

\begin{frontmatter}

\title{Non-local energetics of random heterogeneous lattices}

\author[ctu]{Jan Zeman\corref{cor}}
\ead{zemanj@cml.fsv.cvut.cz}
\ead[url]{http://mech.fsv.cvut.cz/\~{}zemanj}
\cortext[cor]{Corresponding author. Tel.:~+420-2-2435-4482;
fax~+420-2-2431-0775}
\author[tue]{Ron H.J. Peerlings}
\ead{R.H.J.Peerlings@tue.nl}
\ead[url]{http://www.mate.tue.nl}
\author[tue]{Marc G.D. Geers}
\ead{M.G.D.Geers@tue.nl}
\ead[url]{http://www.mate.tue.nl}
\address[ctu]{Department of Mechanics, Faculty of Civil
  Engineering, Czech Technical University in Prague,
  Th\' akurova 7, 166 29 Prague 6, Czech Republic}

\address[tue]{Eindhoven University of Technology, Department of
  Mechanical Engineering, Materials Technology, PO Box 513, 5600 MB Eindhoven, The Netherlands}

\begin{abstract}
In this paper, we study the mechanics of statistically non-uniform two-phase
elastic discrete structures. In particular, following the methodology proposed
in~(Luciano and Willis, \emph{Journal of the
Mechanics and Physics of Solids} 53, 1505--1522, 2005),
energetic bounds and estimates of the Hashin-Shtrikman-Willis type are developed for
discrete systems with a heterogeneity distribution
quantified by second-order spatial statistics. As illustrated by
three numerical case studies, the resulting expressions for
the ensemble average of the potential energy are
fully explicit, computationally feasible and free of adjustable parameters. Moreover, the
comparison with reference Monte-Carlo simulations confirms a notable improvement
in accuracy with respect to 
approaches based solely on the first-order statistics.
\end{abstract}

\begin{keyword}
inhomogeneous material~(B); 
structures~(B); 
energy methods~(C); 
probability and statistics~(C)
\end{keyword}

\end{frontmatter}

\section{Introduction}\label{sec:intro}

Discrete material models, which represent a material as a network of
particles interacting via inter-particle potentials, have received a
steadily increasing attention in the fields of theoretical,
computational and applied materials science in the last decade, see,
e.g., reviews
by~\cite{Ostoja-Starzewski:2002:LMM,Alava:2006:SMF,Blanc:2007:ACL} and
references therein. From the engineering point of view, the interest
has been nourished by the possibility to address, in a conceptually
simple framework, the interplay among the intrinsic material
heterogeneities, discreteness and randomness on different levels of
resolution. Examples of the successful application of discrete models
include the simulation of materials with complex microstructures such
as paper~\citep{Ostoja-Starzewski:2001:RFM,Bronkhorst:2003:MP2D},
biological materials~\citep{Arnoux:2002:NDM}, low-density
materials~\citep{Christensen:2000:MCLD} and granular
media~\citep{Miehe:2004:FMMT}. Another field of application where the
discrete modeling concept plays an irreplaceable role is the analysis
of localized phenomena in heterogeneous media, such as local buckling
and delamination in thin films, e.g.~\citep{Jagla:2007:MDD,
  Vellinga:2008:IBC}, or, most typically, the simulation of damage and
fracture processes in cohesive-frictional materials,
see~\citep{Lilliu:2003:LFM,Ibrahimbegovic:2003:MMDM,Cusatis:2006:CSL,Chudoba:2006:SMMFI,Vorechovsky:2006:SMMFII}
to cite a few.

The closely related problem of establishing a rigorous link between a discrete
representation and its equivalent continuum response has been the focus of
numerous works. The goal of such studies is to identify an appropriate continuum
representation with materials constants directly related to the underlying
discrete system. Within the computational approaches, perhaps the most prominent
example is the 'local' Quasi-Continuum method, introduced
by~\cite{Tadmor:1996:QAD}, in which the effective behavior of a material point
is adaptively deduced from the response of its finite neighborhood, constrained
by the Cauchy-Born rule; see~\cite{Miller:2002:QM} for further details and
discussion on related concepts.

Complementary to the numerical treatment, a notable research effort has been put
into a rigorous interpretation of infinite-size limits of discrete models from
the point of view of standard and generalized continua. For systems interacting
via potentials satisfying suitable growth conditions, a general local continuum
representation is currently available, proven using the tools of
$\Gamma$-convergence~\citep{Alicandro:2005:GIR} or the thermodynamical limit
procedure due to~\cite{Blanc:2002:FMM}. These results were further utilized to
provide rigorous bounds on the effective macroscopic conductivity of discrete
lattices~\citep{Braides:2004:BEB} or as a theoretical support for the
quasi-continuum approximation to \emph{ab initio} calculations of material
constants~\citep{Gavinia:2007:QCOF}. Recently, both frameworks were successfully
extended to the stochastic setting, see~\cite{Alicandro:2007:MDR,alicicglo09}
and~\cite{Blanc:2007:ESM,Blanc:2007:SHR}. In addition, the
validity of the Cauchy-Born rule were rigorously
examined in~\cite{Friesecke:2002:VFCB} and \cite{Berezhnyy:2006:CL3D} for both
regular and irregular networks, thus explicitly demonstrating  
potential
limitations of the Cauchy-Born type continuum approximation when applied to
discrete localized phenomena.

In such cases, generalized continuum theories provide a well-established way to
introduce an internal lengthscale to the problem, thereby preventing the
pathological localization or singularities of mechanical fields, see e.g.
reviews~\citep{Ganghoffer:1999:RNM,Bazant:2002:NIF,Eringen:2002:NCFT}.
Particular examples of gradient-based theories include the one- and
two-dimensional large strain elasticity studies
by~\cite{Triantafyllidis:1993:HOG} and~\cite{Bardenhagen:1994:DHO}, the
micropolar continuum description of~\cite{Pradel:1998:CMEL}
and~\cite{Martinsson:2007:HMP} or the arbitrary-order convex expansion scheme
due to~\cite{Arndt:2005:DHOG}. Finally, an exhaustive analysis of
one-dimensional systems with generic nearest-neighbor interactions rigorously
demonstrated that the limit behavior may exhibit both diffuse as well as
localized cracking, in the deterministic~\citep{Braides:2002:CLDS} and
stochastic~\citep{Iosifescu:2001:VFD} setting, including the numerical
analysis of discrete-to-continuum coupling in the deterministic
case~\citep{Blanc:2005:APM}. More recently, these results were extended
by~\cite{Braides:2008:OPDM}, who studied continuum limits of lattices with
randomly distributed defects and showed that the effective behavior is governed
by percolation phenomena. The treatment of finite-size discrete systems
is, to our best knowledge, much less developed and is typically limited to fitting of
phenomenological constitutive relations to numerical simulation results,
cf.~\citep[and references therein]{Rinaldi:2007:SDT,Grassl:2010:MSA}.

In the present paper, we address in detail a specific problem of the mechanics
of random discrete media, namely the formulation of total
potential energy estimates for finite binary lattices with a fixed geometry and
a heterogeneity distribution described by
second-order spatial statistics. Variational bounds and estimates are
established by extending the recent works of
\cite{Luciano:2005:FE,Luciano:2006:HSB} related to the Galerkin discretization
of the stochastic Hashin-Shtrikman-Willis~($\HSW$) variational
principles~\citep{Hashin:1962:OSVP,Willis:1977:BSC}. Our motivation for focusing
on finite-sized systems and the potential energy
instead of the more common continuum setting and local stress- or strain-related
quantities arises from the following considerations:

\begin{itemize}

\item The separation-of-scales 
assumption is inherently not applicable when dealing with
finite discrete structures. This renders the resulting
theory well-suited to predict the statistics of localized responses.

\item In view of recent advances in variational models of complete
  damage~\citep{Bouchitte:2008:CDP,Mielke:2009:CDE,Mielke:2009:CDEb}, the global
  energetic bounds/estimates provide an essential ingredient for the development
  of 'rational' damage mechanics of discrete
  networks.

\item Due to the simple structure of the underlying theory, the
  relevant statistics can be characterized with a generality
  which is currently not available for continuous systems,
  cf.~\Sref{randomization}.

\item Highly accurate estimates of the quantities of interest for
  general structures and loading regimes can be determined on the
  basis of simple Monte-Carlo simulations, see also
  \cite{Sharif-Khodaei:2008:MBM} for a related one-dimensional study
  in the continuous setting.

\end{itemize}

The remainder of the paper is organized as follows. In \Sref{setting},
the relevant steps of the problem definition are specified for both
deterministic and randomized systems. Energetic bounds and estimates
are derived in \Sref{HSW}. In \Sref{num_examples}, results of
numerical studies are presented to assess their accuracy and
limitations. Finally, \Sref{concl} collects concluding
remarks and comments on future extensions of the method.

\section{Problem description}\label{sec:setting}
This Section is devoted to the problem statement, starting with a
brief summary of structural statics, in order to introduce our
notation, followed by the specification of the stochastic framework
and quantities of interest.  Standard notations and results of linear
algebra are employed~\citep{Horn:1990:MA}, with $a$, $\vek{a}$ and
$\mtrx{A}$ denoting a scalar quantity, a vector~(column matrix) and a
generic matrix, respectively. Matrix indexing is used when
appropriate, i.e. given two index sets $\vek{i}$ and $\vek{j}$ with
cardinalities $\card{\vek{i}}$ and $\card{\vek{j}}$,
$^{\vek{i}\vek{j}}\mtrx{A} \in
\set{R}^{\card{\vek{i}}\times\card{\vek{j}}}$ denotes the appropriate
sub-matrix of $\mtrx{A}$, while $^{\vek{i}:}\mtrx{A}$ gives the
corresponding matrix rows and $^{\vek{i}:}\mtrx{A}\trn$ abbreviates
$^{\vek{i}:}( \mtrx{A}\trn )$. Moreover, the matrix formalism developed
in~\cite{Jirasek:2001:IAS} for general discrete structures is
systematically adopted.

\subsection{Summary of discrete media mechanics}\label{sec:summ_discr_med}
Consider a discrete structure consisting of $\numno$ nodes with
coordinates $\vek{x}_i \in \set{R}^d$, $i = \{1,2, \ldots, \numno \}$
and~$d \in \{2,3\}$, connected by $\numel$ discrete elements. On the
level of a single element $e \in \{1,2,\ldots,\numel\}$, the
generalized kinematic equations take the form
\begin{equation}\label{eq:kM}
\eV\fel = \kM\fel \dV\fel,
\end{equation}
where $\eV\fel \in \set{R}^\numstress$ is the vector of generalized
strains, the vector $\dV\fel \in \set{R}^{2\numdof}$ contains the
$\numdof$ generalized displacements at both element nodes and $\kM\fel
\in \set{R}^{\numstress \times 2\numdof}$ denotes the element
kinematic matrix. The corresponding generalized element stresses
$\sV\fel \in \set{R}^\numstress$ then follow from
\begin{equation}\label{eq:mM}
\sV\fel = \mM\fel \eV\fel,
\end{equation}  
with $\mM\fel \in \set{R}^{\numstress \times \numstress}$ denoting a
matrix of generalized material stiffness. On the structural level, the
relations~\eqref{kM} and~\eqref{mM} can be assembled into the form
\begin{eqnarray*}
\eV = \kM \dV, &&
\sV = \mM \eV,
\end{eqnarray*}
where, e.g., $\kM \in \set{R}^{\numel\numstress \times
  \numno\numdof}$, $\mM \in \set{R}^{\numel\numstress \times
  \numel\numstress}$ and $\dV \in \set{R}^{\numno\numdof}$ stand for
the global kinematic matrix, block-diagonal generalized stiffness
matrix and displacement vector defined as
\begin{eqnarray}
\kM = \assembly_{\el=1}^{\numel} \kM\fel, 
& \displaystyle
\mM = \assembly_{\el=1}^{\numel} \mM\fel
&
\mbox{and~~}
\dV = \assembly_{\el=1}^{\numel} \dV\fel,
\end{eqnarray}
with the symbol $\assembly$ representing the assembly operation,
cf.~\citep[Appendix A]{Jirasek:2001:IAS}. The remaining matrices and vectors are
defined analogously.

In order to specify kinematic constraints on the structure, we partition the
problem degrees of freedom~(\DOFs) into two sets
\begin{equation}\label{eq:fix_free}
\fixed \cup \free = \{1, 2, \ldots, \numno \numdof \}, 
\quad
\fixed \cap \free = \emptyset,
\quad
\mathrm{ker}\left( ^{:\free}\kM \right) = \{ \vek{0} \},
\end{equation}
where $\fixed$ and $\free$ collect the known~(constrained) and unknown
\DOFs~and $\mathrm{ker}( \mtrx{A} )$ denotes the kernel of a matrix
$\mtrx{A}$. Note that the last condition in~\Eref{fix_free} enforces
the elimination of rigid-body modes. When subjecting the structure to
an additional nodal load $^{\free}\fV \in \set{R}^{\card{\free}}$
acting on free~\DOFs, the unknown displacements $^{\free}\dV$ can be
found by solving the unconstrained quadratic optimization problem
\begin{equation}\label{eq:optim_problem}
^{\free}\dV
= 
\arg \min_{\trial{\dV} \in \set{R}^{\card{\free}}}
\EF( \trial{\dV} ),
\end{equation}
where the total potential energy function $\EF :
\set{R}^{\card{\free}} \rightarrow \set{R}$ is provided by
\begin{equation}\label{eq:det_energy}
\EF( \trial{\dV} ) 
=
\half
\myMatrix{cc}{%
\trial{\dV}\trn & ^{\fixed}\dV\trn
}
\myMatrix{cc}{%
^{\free\free}\stM & ^{\free\fixed}\stM \\
^{\fixed\free}\stM & ^{\fixed\fixed}\stM
}
\myMatrix{c}{%
\trial{\dV} \\ ^{\fixed}\dV
}
-
\trial{\dV}\trn\,{}^{\free}\fV,
\end{equation}
with, for example, $^{\free\fixed}\stM={}^{\free:} \sM
\mM~^{:\fixed}\kM \in \set{R}^{\card{\free}\times\card{\fixed}}$ being
a sub-matrix of the global stiffness matrix $\stM = \sM \mM \kM$. The
symbol ``$\arg\min$'' appearing in \Eref{optim_problem} denotes the
minimizer of the objective function verifying
\begin{equation}
\EF( ^{\free}\dV ) \leq \EF( \trial{\dV} ) 
\quad 
\forall \trial{\dV} \in \set{R}^{\card{\free}},
\end{equation}
where the equality is attained only for the test displacement
$\trial{\dV}$ coinciding with the true solution due to positive
definiteness of $^{\free\free}\stM$. The optimality conditions for
$^{\free}\dV$ then yield the global equilibrium equations in the form
\begin{equation}\label{eq:equil_eqs}
{}^{\free\free}\stM\,{}^{\free}\dV
=
{}^{\free}\fV 
- 
{}^{\free\fixed}\stM\,{}^{\fixed}\dV.
\end{equation} 

\subsection{Stochastic setting}\label{sec:randomization}

We now proceed with the introduction of a suitable framework for
binary random discrete media, i.e. structures in which every element
can be found in one of two distinct states $r \in \{1,2\}$. Due to the
discrete nature of the problem at hand, the ensemble space~$\enspc$
collecting all structural configurations is finite-dimensional and as
such can be enumerated using an index $\rl$,
\begin{equation}
\rl \in \enspc = \left\{ 1, 2, \ldots, 2^\numel \right\}.
\end{equation}
The complete statistical characterization of the discrete stochastic
system is then simply provided by assigning probabilities $\pd(\rl)$
to individual configurations $\rl$ stored in the probability
distribution vector
\begin{equation}
\pdV \in \smplx 
= 
\left\{ 
\trial{\pdV} \in \set{R}^{\card{\enspc}},
\trial{\pd}(\rl) \geq 0 \ \forall \rl \in \enspc, 
\sum_{\alpha=1}^{\card{\enspc}} \trial{\pd}(\rl) = 1
\right\}.
\end{equation}
The ensemble average of a configuration-dependent quantity $f(\rl)$
for a given probability distribution $\pdV \in \smplx$ is defined as
\begin{equation}
\avg{f}{\pdV}
=
\sum_{\rl=1}^{\card{\enspc}}
f(\rl) \pd(\rl).
\end{equation}

Of particular importance is the state characteristic vector $\cV\phs{r}(\rl)$
defined via
\begin{equation}
\cF\fel\phs{r}(\rl) 
=
\left\{
 \begin{array}{cl}
 1 & \mbox{if element $\el$ is in state $r$ for configuration $\rl$,}\\
 0 & \mbox{otherwise,}
 \end{array}
\right.
\end{equation}
quantifying the spatial distribution of individual states in a given
configuration $\rl$. For the current, binary, case, $\cV\phs{r}(\rl)$
can be explicitly expressed in the form
\begin{eqnarray}\label{eq:char_func_def}
\cV\phs{1}(\rl) = (\rl-1)_{\set{B}},
&&
\cV\phs{1}(\rl) + \cV\phs{2}(\rl) = \vek{1},
\end{eqnarray}
where $n_{\set{B}}$ provides the value of a natural number $n$ in the
binary notation using $\numel$~digits arranged in a column matrix,
see~\Fref{ens_spc} for an illustration.

\begin{figure}[b]
 \centering
 \includegraphics{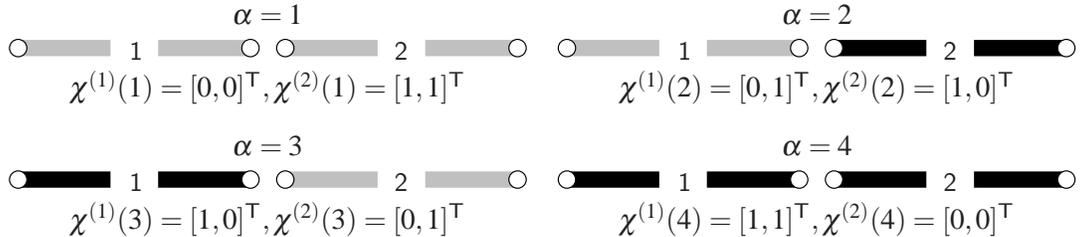}
\caption{Ensemble space and state characteristic vectors for a
  two-element structure; state $r=1$ corresponds to a black element,
  $r=2$ to a gray element.}
\label{fig:ens_spc}
\end{figure}

Following the analogy with the quantification of the spatial
statistics of random heterogeneous media,
e.g.~\citep{Torquato:2002:RHM}, we introduce an $\numel \times \numel$
two-unit probability matrix related to a given probability
distribution $\pdV$ in the form
\begin{equation}\label{eq:two_element_def}
\PM\phs{rs}
=
\avg{\cV\phs{r}\cV\phs{s}\trn}{\pdV}
=
\sum_{\rl=1}^{2^\numel}
\cV\phs{r}(\rl) \cV\phs{s}(\rl)\trn \pd(\rl),
\end{equation}
where an individual entry $P_{e e'}\phs{rs}$ represents the probability
of states $r$ and $s$ being assigned to elements $e$ and $e'$~(note
that the explicit dependence on $\pdV$ is dropped for the sake of
notational brevity). The two-unit matrices related to distinct states
are not independent of each other: using~\eqref{char_func_def}$_2$
yields
\begin{eqnarray}
\PM\phs{12} & = & \pV\phs{1} \vek{1}\trn -\PM\phs{11}, \label{eq:PM_12} \\
\PM\phs{21} & = & \vek{1} \pV\phs{1}\trn - \PM\phs{11}, \label{eq:PM_21} \\
\PM\phs{22} & = & 
\vek{1}\vek{1}\trn - \pV\phs{1} \vek{1}\trn - 
\vek{1} \pV\phs{1}\trn + \PM\phs{11},
\end{eqnarray}
with $\pV\phs{r} = \diag{\PM\phs{rr}}$. It is therefore sufficient to
concentrate on the statistics $\PM\phs{11}$ below.

It directly follows from the definition~\eqref{two_element_def} that
any two-element probability matrix has to be located in a convex hull
of rank-one products of the characteristic vectors of individual
configurations:
\begin{equation}
\PM\phs{11} 
\in \bqp 
=
\mathrm{conv} \left\{ 
 \cV\phs{1}(1) \cV\phs{1}(1)\trn, \cV\phs{1}(2) \cV\phs{1}(2)\trn,  
 \ldots, \cV\phs{1}(\card{\enspc}) \cV\phs{1}(\card{\enspc})\trn
\right\},
\end{equation}
coinciding with the Boolean quadratic polytope completely
characterized by~\cite{Padberg:1989:BQP}. Conversely, two-element
probability matrices can be used as a convenient re-parameterization
of $\smplx$. To that end, we introduce a set storing all probability
distributions compatible with matrix $\PM\phs{11}$ as
\begin{equation}
\admPD( \PM\phs{11} ) = \left\{ \trial{\pdV} \in \smplx, \PM\phs{11} =
\avg{\cV\phs{1} \cV\phs{1}\trn}{\trial{\pdV}} \right\},
\end{equation}
thus establishing a partial statistical characterization when full
information is not available.\footnote{%
To the best of our knowledge, no such results are available for
general multi-unit probability functions. In addition, the treatment of
higher-order statistics leads to a substantial increase of storage and
computing requirements. Therefore, we limit our attention to the second-order
framework and leave its extension to future work.}

At this point, we can introduce the terminology used hereafter. Given a
two-unit probability matrix $\PM\phs{11}$, the associated discrete system is
called \emph{statistically uniform} if
\begin{eqnarray}\label{eq:def_stat_uniform}
p\phs{1}_e = p, &
P\phs{11}_{e e'} = P( \vek{x}_e - \vek{x}_e' ) 
& \mbox{for } e, e' = 1, 2, \ldots, \numel,
\end{eqnarray}
i.e. the one-unit
probabilities of individual elements are independent of $e$, while  the two-unit
probabilities depend only on the difference of coordinates of element centers
$\vek{x}_e$ and $\vek{x}_e'$. In addition, a function $f( \PM\phs{11} )$ is \emph{statistically local}~(or
first-order) if it depends on the first-order statistics only:
\begin{eqnarray}\label{eq:def_stat_local}
f( \PM\phs{11} ) = g( \pV\phs{1} ).
\end{eqnarray}
In the opposite case, we use the adjectives \emph{statistically non-local}~(or
second-order).

It is perhaps instructive to briefly comment on similarities and differences
with the analogous development in the continuous case. In particular, the
two-element probability matrix is a direct analogue of the two-point probability
functions for generic statistically non-uniform and anisotropic
microstructures~\citep{Willis:1981:VRM,Torquato:2002:RHM}. The set $\bqp$ then
coincides with the set of all admissible two-point probability functions, the analysis of which
has recently received substantial attention in the field of random heterogeneous
materials, eventually resulting in a complete characterization for statistically
uniform media as shown by~\cite{Quintanilla:2008:NSC}; see
also~\cite{Jiao:2007:MTV} for further discussion. In this connection, the
advantage of considering finite discrete systems
becomes immediately visible, since the characterization
of~\cite{Padberg:1989:BQP} is free of assumption of statistical
uniformity. 

\subsection{Statics of two-phase random lattices}
When assigning specific material properties to each state $r$, the previously
introduced framework can be readily adopted to two-phase elastic heterogeneous
lattices. In particular, the configuration-dependent material stiffness matrix
on the element level, $\mM\fel(\rl)$, can be written as
\begin{equation}\label{eq:mtrl_stiff_mtrx}
\mM\fel(\rl) = \sum_{r=1}^{2} \cF\fel\phs{r}(\rl) \mM\fel\phs{r},
\end{equation}
where $\mM\fel\phs{r}$ are generalized material stiffness matrices of
individual states. On the structural level, the stiffness
distribution is described by
\begin{equation}
\mM( \rl ) 
=
\assembly_{\el=1}^{\numel} \mM\fel(\rl)
=
\sum_{r=1}^2 
\left( 
 \assembly_{\el=1}^{\numel} 
 \cF\fel\phs{r}(\rl) \mM\fel\phs{r}
\right)
=
\sum_{r=1}^2 
\cV\phs{r}(\rl) \sprod \mM\phs{r},
\end{equation}
where $\vek{a} \sprod \mtrx{A}$ denotes a block Hadamard-like product
implementing the assembly operation. 

Consider now the response of a discrete structure with a
\emph{stochastic} configuration-dependent generalized material stiffness matrix
$\mM(\rl)$ subject to \emph{deterministic} loading conditions specified in terms
of prescribed displacements ${}^{\fixed}\dV$ and generalized nodal forces
$^{\free}\fV$. For each configuration $\rl \in \enspc$, the energy minimizer is
defined as
\begin{equation}\label{eq:minizer_conf}
^{\free}\dV( \rl ) 
=
\arg \min_{\trial{\dV} \in \set{R}^{\card{\free}}}
\EF( \trial{\dV}; \rl ),
\end{equation}
with the stored energy function introduced analogously
to~\eqref{det_energy}:
\begin{equation}\label{eq:stoch_energy}
\EF( \trial{\dV}; \rl ) 
=
\half
\myMatrix{cc}{%
\trial{\dV}\trn & {}^{\fixed}\dV\trn
}
\myMatrix{cc}{%
^{\free\free}\stM( \rl ) & ^{\free\fixed}\stM( \rl ) \\
^{\fixed\free}\stM( \rl ) & ^{\fixed\fixed}\stM( \rl )
}
\myMatrix{c}{%
\trial{\dV} \\ ^{\fixed}\dV
}
-
\trial{\dV}\trn~^{\free}\fV.
\end{equation}
The ensemble average of the optimal energy for a given probability
distribution $\pdV$ is then simply a weighted sum
\begin{equation}\label{eq:avg_energy_min}
\avg{\EF(^{\free}\dV)}{\pdV} 
=
\sum_{\rl=1}^{\card{\enspc}}
\EF\left( ^{\free}\dV( \rl ) \right)
\pd(\rl).
\end{equation}

A full specification of the probability distribution is, however,
rarely available and even with complete information at hand,
evaluating~\eqref{avg_energy_min} requires the solution of $2^\numel$
problems, which becomes rapidly unfeasible even for moderate-size
problems. Therefore, we rely on a partial statistical characterization
in terms of two-element probabilities and attempt to establish
energetic bounds/estimates in the form
\begin{equation}\label{eq:two_sided_estimates}
\sHF\low( \PM\phs{11} ) 
 \leq 
 \sHF\appr( \PM\phs{11} ) \approx
 \avg{\EF(^{\free}\dV)}{\trial{\pdV}} 
 \leq
\sHF\up( \PM\phs{11} ) 
\quad
\forall \trial{\pdV} \in \admPD( \PM\phs{11} ),
\end{equation}
reflecting the limited probabilistic characterization. In addition to
the energetics, we also provide elementary statistics of the nodal
displacements related to the bounds and estimates of the energy
introduced in~\Eref{two_sided_estimates}.

\section{Hashin-Shtrikman-Willis-type estimates}\label{sec:HSW}
%
In this Section, we provide explicit energetic bounds for random
finite-size networks by reconsidering the Hashin-Shtrikman variational
principles for random heterogeneous media in the discrete setting. To
make the exposition more readable, the derivations are structured in
six consecutive steps.

\subsection{Reference structure and generalized polarization stresses}
%
Following the conceptual lead of~\cite{Hashin:1962:OSVP}, we introduce
a reference deterministic structure characterized by a
positive-definite generalized material stiffness matrix $\mM\phs{0}$
and consider a realization-dependent quadratic form
\begin{equation}\label{eq:comp_form}
\half
\myMatrix{cc}{%
\trial{\tV}\trn & \trial{\eV}\trn
}
\myMatrix{cc}{%
\left( \mM\phs{0} - \mM( \rl ) \right)\inv & \uM \\
\uM & \left( \mM\phs{0} - \mM( \rl ) \right)
}
\myMatrix{c}{%
\trial{\tV} \\
\trial{\eV}
},
\end{equation}
where $\trial{\eV} \in \set{R}^{\numel\numstress}$ is a test
generalized strain vector and the auxiliary variable $\trial{\tV} \in
\set{R}^{\numel\numstress}$ will be commented on later. By virtue of
the Schur complement lemma, cf.~\cite[Section~7.7.6]{Horn:1990:MA},
the form~\eqref{comp_form} is positive-semidefinite as long as $(
\mM\phs{0} - \mM( \rl ))$ is positive definite, leading to a bound
\begin{equation}\label{eq:HS_estimate}
\half
\trial{\eV}\trn \mM(\rl) \trial{\eV}
\leq
\half
\trial{\eV}\trn \mM\phs{0} \trial{\eV}
+
\trial{\tV}\trn \trial{\eV}
+
\half
\trial{\tV}\trn ( \mM\phs{0} - \mM(\rl) )\inv \trial{\tV}
,
\end{equation}
with the equality reserved for $\trial{\tV}$ equal to
\begin{equation}\label{eq:HS_estimate_1}
\sopt{\tV}( \rl ) 
=
- \left( \mM\phs{0} - \mM( \rl ) \right) \trial{\eV} 
=
\trial{\sV}( \rl ) - \mM\phs{0} \trial{\eV},
\end{equation}
which minimizes the right hand side of~\eqref{HS_estimate} for a given
realization $\rl$ and test generalized strain field $\trial{\eV}$. The
variable $\sopt{\tV}$ therefore corresponds to a \emph{generalized
  polarization stress} associated with the reference stiffness matrix
$\mM\phs{0}$ and generalized
strain~$\trial{\eV}$~\citep{Hashin:1962:OSVP}.

Inequality~\eqref{HS_estimate} leads, for an arbitrary
positive-definite $(\mM\phs{0}-\mM(\rl))$, to an upper bound on the
configuration-dependent internal energy of the structure expressed in
terms of an auxiliary structure with stiffness $\mM\phs{0}$ and
identical topology, which is subject to generalized strain
$\trial{\eV}$ and polarization stress $\trial{\tV}$. Moreover, when
the polarization stress is compatible with the difference in stiffness
distribution between the two configurations $\mM\phs{0}$ and
$\mM(\rl)$ and the imposed generalized strain, the gap between and
upper bound and the true value vanishes.

Notice that when the optimization with respect to $\trial{\tV}$ is
performed exactly, the resulting equality~\eqref{HS_estimate} is
independent of the choice of the reference structure. Otherwise, e.g.
when the set of admissible generalized polarizations $\trial{\tV}$ is
constrained, different choices of $\mM\phs{0}$ generate different
upper bounds on the stored energy. The most restrictive upper bound
then corresponds to $\mM\phs{0}$ chosen as close as possible to the
actual generalized stiffness $\mM(\rl)$ while maintaining the positive
definiteness of $(\mM\phs{0} - \mM(\rl) )$.\footnote{%
Observe from~\Eref{HS_estimate_1} that for $\mM\phs{0} \rightarrow
\mM(\rl)$, $\sopt{\tV} \rightarrow \vek{0}$ for an arbitrary
$\trial{\eV}$ and equality in~\eqref{HS_estimate} is recovered.}
To make the choice of the reference structure independent of $\rl$, it
follows from the specific form of matrices $\mM( \rl )$
in~\eqref{mtrl_stiff_mtrx} that the ordered eigenvalues of optimal
$\mM\phs{0}$ are determined as the minima of the corresponding
eigenvalues of the individual states:
\begin{eqnarray*}
\lambda\phs{0}_i = \min_r \lambda\phs{r}_i &&
i = 1, 2, \ldots, \numel\numstress,
\end{eqnarray*}
where $\lambda\phs{r}_i$ corresponds to the $i$-th ordered eigenvalue
of matrix $\mM\phs{r}$, cf.~\citep[Section~4]{Dvorak:1999:NEOP}. 

A completely analogous procedure can be executed when selecting the
reference structure such that $( \mM\phs{0} - \mM(\rl))$ becomes
negative-definite, leading to a lower bound on the stored energy. For
an indefinite $( \mM\phs{0} - \mM(\rl) )$, a variational estimate of
the stored energy is obtained, see e.g.~\cite{Luciano:2006:HSB} for
additional discussion.

\subsection{Discrete Hashin-Shtrikman variational principles}

Once inequality~\eqref{HS_estimate} has been established, the
Hashin-Shtrikman variational principles directly follow from the
original energy minimization problem~\eqref{minizer_conf} for a
configuration $\alpha$. Assume that $( \mM\phs{0} - \mM( \rl ))$ is
positive definite and introduce a kinematically admissible generalized
strain $\trial{\eV} \in \set{R}^{\numel\numstress}$ obtained from a
test displacement $\trial{\dV} \in \set{R}^{\card{\free}}$ via
\begin{equation}\label{eq:HS_ref_strain}
\trial{\eV}
=
\myMatrix{cc}{%
{}^{:\free}\kM & {}^{:\fixed}\kM
}
\myMatrix{c}{%
\trial{\dV} \\
{}^{\fixed}\dV
}.
\end{equation}
Then, we obtain the upper bound
\begin{eqnarray}
\EF( \trial{\dV}; \rl ) 
& \by{stoch_energy}{=} & 
\half 
\trial{\eV}\trn \mM( \rl ) \trial{\eV} - \trial{\dV}\trn{}^{\free}\fV 
\by{HS_estimate}{\leq}
\half \trial{\eV}\trn \mM\phs{0} \trial{\eV} - \trial{\dV}\trn {}^{\free}\fV
\label{eq:HS_deter_energy}
\\
& 
+ 
& 
\trial{\tV}\trn \trial{\eV}
+
\half
\trial{\tV}\trn ( \mM\phs{0} - \mM( \rl ) )\inv \trial{\tV}
, \nonumber
\end{eqnarray}
which yields a variational characterization of the true
displacement-polarization pair in the form
\begin{equation}\label{eq:hs_problem_full}
\left( ^{\free}\dV(\rl), \tV(\rl) \right)
=
\arg 
\min_{\trial{\dV} \in \set{R}^{\card{\free}} }
\min_{\trial{\tV} \in \set{R}^{\numel \numstress}}
\UF( \trial{\dV}, \trial{\tV}; \rl ), 
\end{equation}
where the Hashin-Shtrikman energy function $\UF$ is defined as
\begin{equation}\label{eq:HS_energy_func_def}
\UF( \trial{\dV}, \trial{\tV}; \rl ) 
=
\EF\phs{0}( \trial{\dV} )
+
\trial{\tV}\trn \trial{\eV}
+
\half
\trial{\tV}\trn ( \mM\phs{0} - \mM( \rl ) )\inv \trial{\tV},
\end{equation}
with $\EF\phs{0}$ denoting the potential energy of the reference structure~(as
introduced in~\Eref{stoch_energy}).

Considering an arbitrary reference media and upon exchanging the order
of optimization, problem~\eqref{hs_problem_full} is extended into its
final form:
\begin{equation}\label{eq:HS_full_2}
\left( {}^{\free}\dV(\rl), \tV(\rl) \right)
=
\arg 
\stat_{\trial{\tV} \in \set{R}^{\numel \numstress}}
\left( 
 \min_{\trial{\dV} \in \set{R}^{\card{\free}} }
 \UF( \trial{\dV}, \trial{\tV}; \rl ) 
\right),
\end{equation}
where the actual meaning of the ``$\stat$'' operation~(minimum,
maximum or critical point) depends on the choice of the reference
structure's stiffness $\mM\phs{0}$.

\subsection{Condensed variational principle}

Similarly to~\cite{Willis:1977:BSC}, we proceed with the condensation
of the kinematic variables from the Hashin-Shtrikman function by
relating the optimal displacement vector $\sopt{\dV}$ to an arbitrary
test polarization stress $\trial{\tV}$. Due to the linearity of the
problem, the actual value can be expressed as a superposition of two
auxiliary solutions
\begin{equation}\label{eq:HS_displ_split}
\myMatrix{c}{%
 ^{\free}\sopt{\dV} \\
 {}^{\fixed}\dV
}
=
\myMatrix{c}{%
 {}^{\free}\dV\cmp{0} \\
 {}^{\fixed}\dV
}
+
\myMatrix{c}{%
 ^{\free}\sopt{\dV}\cmp{1} \\
 \vek{0}
},
\end{equation}
where $^{\free}\dV\cmp{0}$ denotes the polarization-independent
displacement of the reference structure subject to prescribed
displacements $^{\fixed}\dV$ and nodal forces $^{\free}\fV$, while
$^{\free}\sopt{\dV}\cmp{1}$ is the displacement resulting from an
internal generalized polarization stress $\trial{\tV}$ with
$^{\fixed}\dV = \vek{0}$ and $^\free \fV = \vek{0}$. The values of
both components follow from the equilibrium
equations~\eqref{equil_eqs}:
\begin{eqnarray}
^{\free\free}\stM\phs{0}~^{\free}\dV\cmp{0}
& = &
{}^{\free}\fV
-
{}^{\free\fixed}\stM\phs{0}\,{}^{\fixed}\dV,
\label{eq:system_1} \\
{}^{\free\free}\stM\phs{0}
\;{}^{\free}\sopt{\dV}\cmp{1}
& = &
-^{\free:}\sM \trial{\tV}.
\label{eq:system_2}
\end{eqnarray}

After resolving the inner optimization in~\Eref{HS_full_2}, we proceed
with determining the optimal polarization. Introducing the solutions
of~\eqref{system_1} and~\eqref{system_2} into the two-variable
function~\eqref{HS_full_2} and exploiting the optimality
conditions~\eqref{system_1} and~\eqref{system_2} yields, after some
algebraic manipulations discussed in detail
in~Appendix~\ref{sec:cond_energy}, the characterization of the optimal
generalized polarization stresses in the form:
\begin{equation}\label{eq:rl_HS_otp}
\tV( \rl ) 
= 
\arg \stat_{\trial{\tV} \in \set{R}^{\numel \numstress}}
\HF( \trial{\tV}; \rl ),
\end{equation}
with the condensed energy function expressed as
\begin{equation}\label{eq:cond_energy}
\HF( \trial{\tV}; \rl )
=
\HF\phs{0} 
+
\trial{\tV}\trn\eV\cmp{0}
+
\half
\trial{\tV}\trn 
( \mM\phs{0} - \mM( \rl ) )\inv
\trial{\tV}
-
\half
\trial{\tV}\trn 
\gM\phs{0}
\trial{\tV},
\end{equation}
where $\HF\phs{0} = \EF\phs{0}( {}^{\free}\dV\cmp{0} )$ corresponds to
a stationary value of the potential energy of 
the reference structure, $\eV\cmp{0}$ is the associated generalized strain determined
from~$\dV\cmp{0}$ and~\Eref{HS_ref_strain} and $\gM\phs{0}$ is the
discrete counterpart of the Green function-related non-local
operators introduced by~\cite{Willis:1977:BSC} and~\cite{Luciano:2005:FE},
linking the kinematic quantities to the generalized polarization
stress via
\begin{equation}\label{eq:gamma_0_def}
\sopt{\eV}\cmp{1}
=
{}^{:\free}\kM~^\free\sopt{\dV}\cmp{1}
\by{system_2}{=}
-~^{:\free}\kM 
\left(
{}^{\free\free}\stM\phs{0}
\right)\inv 
{}^{:\free}\kM \trn
\trial{\tV}
=
-\gM\phs{0}\trial{\tV}.
\end{equation}

\subsection{Approximation}

When considering a given probability distribution $\pdV \in \smplx$ of all
possible states, the ensemble average of the stationary energy follows from
\begin{equation}\label{eq:ens_avg_energy}
\avg{\EF(^{\free}\dV(\rl))}{\pdV} 
=
\avg{\HF(\tV(\rl))}{\pdV} 
\lessgrta
\avg{\HF(\trial{\tV}(\rl))}{\pdV},
\end{equation}
where the shorthand notation $\lessgrta$ is used to emphasize that
the actual status of the right hand side depends again on the choice of reference
generalized material stiffness matrix $\mM\phs{0}$. The equality
in~\eqref{ens_avg_energy} remains valid since we have assumed so far that the
optimization problem~\eqref{rl_HS_otp} is resolved exactly for every
configuration~$\rl$. With the second-order description of the stochastic system
at hand, a specific ansatz for the generalized polarization stresses given by
\begin{eqnarray}\label{eq:pol_stress_ansatz}
\tV( \rl ) 
\approx 
\sum_{r=1}^2 \cV\phs{r}( \rl ) \sprod \tV\phs{r},
&&
\trial{\tV}( \rl ) 
\approx 
\sum_{r=1}^2 \cV\phs{r}( \rl ) \sprod \trial{\tV}\phs{r},
\end{eqnarray}
is employed to exploit the available second order
statistics~\eqref{two_element_def} optimally,
cf.~\citep{Willis:1977:BSC}. The $\trial{\tV}\phs{r}$ and $\tV\phs{r}$
in~\Eref{pol_stress_ansatz} are \emph{realization-independent} trial
and ``true'' generalized polarization stresses related to state
$r$. 

Introducing the approximations~\eqref{pol_stress_ansatz} and the
expression for the generalized material stiffness
matrix~\eqref{mtrl_stiff_mtrx} into~\Eref{ens_avg_energy} yields a
variational statement in the form
\begin{eqnarray}
\avg{\EF(^{\free}\dV(\rl))}{\pdV} 
& \lessgrta &
\HF\phs{0} 
+
\sum_{r=1}^2 
\trial{\tV}\phs{r} 
\trn
\left( 
\avg{\cV\phs{r}}{\pdV}
\sprod 
\eV\cmp{0}
\right)
\nonumber \\
& + & 
\half
\sum_{r=1}^{2} \sum_{s=1}^2
\trial{\tV}\phs{r}\trn
\left(
 \avg{\cV\phs{r}\cV\phs{s}\trn}{\pdV} \sprod \left( \mM\phs{0} - \mM\phs{r} \right)\inv
\right)
\trial{\tV}\phs{s}
\nonumber \\
& - &
\half
\sum_{r=1}^2 \sum_{s=1}^2
\trial{\tV}\phs{r}
\left( 
 \avg{\cV\phs{r} \cV\phs{s}\trn}{\pdV} \sprod \gM\phs{0}
\right)
\trial{\tV}\phs{s}. \label{eq:HS_two_unit_1}
\end{eqnarray}
Imposing the constraint $\pdV \in \admPD( \PM\phs{11} )$, we can
explicitly evaluate the ensemble averages appearing at the right hand
side of~\Eref{HS_two_unit_1}. Observing that
$$
\PM\phs{rs} \sprod \left( \mM\phs{0} - \mM\phs{r} \right)\inv =
\mtrx{0} \mbox{ for } r \neq s,
$$
as $(\mM\phs{0}-\mM\phs{r})$ is block-diagonal while $\PM\phs{rs}$ is
zero-diagonal for $r \neq s$ due Eqs.~(\ref{eq:PM_12},\ref{eq:PM_21}),
we express the bounds/estimates in terms of the two-unit statistics
\begin{eqnarray}
\avg{\EF(^{\free}\dV(\rl))}{\pdV} 
& \lessgrta &
\HF\phs{0} 
+
\half
\sum_{r=1}^2
\trial{\tV}\phs{r}\trn
\left[ 
 2 \pV\phs{r} 
  \sprod
 \eV\cmp{0}
+
 \left(
  \PM\phs{rr}
   \sprod
  \left( \mM\phs{0} - \mM\phs{r} \right)\inv
 \right)
 \trial{\tV}\phs{r} 
\right.
\nonumber \\
& - & 
\left.
 \sum_{s=1}^2
  \left( 
   \PM\phs{rs} \sprod \gM\phs{0}
  \right)
\trial{\tV}\phs{s}
\right]. \label{eq:HS_two_unit_2}
\end{eqnarray}

\subsection{Non-local energetic bounds and estimates}

The next step of the derivation involves the determination of the
optimal state polarization stresses. The stationary conditions for
function~\eqref{HS_two_unit_2} with respect to $\trial{\tV}\phs{r}$
variables yield a system of linear equations
\begin{equation}\label{eq:HS_phase_optim}
-\left( 
\PM\phs{rr} \sprod\left( \mM\phs{0} - \mM\phs{r} \right)\inv
\right)
\tV_{\pma}\phs{r} 
+ 
\sum_{s=1}^2 
\PM\phs{rs} \sprod \gM\phs{0}
\tV_{\pma}\phs{s} 
=
\pV\phs{r} \sprod \eV\cmp{0}
\end{equation}
to be satisfied by the true state polarization fields $\tV\phs{r}$,
with the subscript $\pma$ again referring to the actual status of
the~$\HSW$ energy function. Employing the optimality
conditions~\eqref{HS_phase_optim} when evaluating the terms in the
square brackets in~\Eref{HS_two_unit_2} provides the desired energetic
bounds and estimates:
\begin{equation}
\sHF_{\pma}( \PM\phs{11} )
=
\HF\phs{0} 
+
\half
\sum_{r=1}^2
\tV_{\pma}\phs{r}
\trn
\left( 
\pV\phs{r}
\sprod 
\eV\cmp{0}
\right),
\end{equation}
with the implicit dependence of $\tV_{\pma}\phs{r}$ on $\PM\phs{11}$
provided by~\Eref{HS_phase_optim}. The final expression consists of
the deterministic value related to the reference problem and the
stochastic contribution, which is \emph{statistically non-local} due
to the incorporation of the discrete Green function~\eqref{gamma_0_def} and
$\PM\phs{11}$ in~\Eref{HS_phase_optim}.

\subsection{Statistics of the response}
%
The known values of state generalized polarization stresses allow us
to obtain elementary statistics of additional quantities apart from
the energy. For example, the mean value of the nodal displacements for
an arbitrary $\pdV \in \admPD( \PM\phs{11} )$ follows from
\begin{eqnarray}
\avg{^{\free}\dV_{\pma}}{\pdV}
& \by{HS_displ_split}{=} &
{}^{\free}\dV\cmp{0}
+
\avg{%
 {}^{\free}\sopt{\dV}_{\pma}\cmp{1}
}{\pdV}
\nonumber \\
& \by{system_2}{=} &
{}^{\free}\dV\cmp{0}
-
\avg{%
\left( 
{}^{\free\free} 
\stM\phs{0} 
\right)\inv
{}^{\free :}\sM
\left( \sum_{r=1}^{2} \cV\phs{r} \sprod \tV\phs{r}_{\pma} \right)
}{\pdV}
\nonumber \\
& = &
{}^{\free}\dV\cmp{0}
-
\sum_{r=1}^2
\left( 
{}^{\free\free} 
\stM\phs{0} 
\right)\inv
{}^{\free :}\sM
\left(
 \pV\phs{r}
   \sprod 
 \tV\phs{r}_{\pma}
\right)
.\label{eq:hs_mean_displ}
\end{eqnarray}

It is worth noting that the previous relations are analogous to the
results derived by~\cite{Luciano:2005:FE,Luciano:2006:HSB} for
\FE-based discretization of the $\HSW$~principles. Therefore, adopting
a similar procedure, means or conditional means of selected local
variables can be established by post-processing the optimal state
polarization stresses. As our focus is on the total
potential energy, we omit explicit expressions for these statistics and refer
the interested reader to~\cite{Luciano:2005:FE,Luciano:2006:HSB} for details.

\section{Examples}\label{sec:num_examples}

Although the theory presented in the previous sections is applicable
to generic linear discrete structures, basic features of the method
are illustrated below for planar truss systems only. Within this
context, the generalized displacement vector and the material and
kinematic matrices introduced in~\Sref{summ_discr_med} specialize
to~\citep{Jirasek:2001:IAS}:
\begin{eqnarray}
\dV\fel 
& = &
\myMatrix{cccc}{%
u_{1,\el} &
v_{1,\el} &
u_{2,\el} &
v_{2,\el}
}\trn,
\quad
\mM\fel
=
\myMatrix{c}{%
\displaystyle \frac{E\fel A\fel}{\ell\fel}}
\nonumber \\
\kM\fel
& = &
\myMatrix{cccc}{%
x_{1,\el} - x_{2,\el} & y_{1,\el} - y_{2,\el} &
x_{2,\el} - x_{1,\el} & y_{2,\el} - y_{1,\el}},
\end{eqnarray}
where, in accordance with~\Fref{element_scheme}, $x_{i,e}$ and
$y_{i,e}$ denote the coordinates of the $e$-th element's nodes,
$u_{i,e}$ and $v_{i,e}$ are the corresponding displacements,
$\ell\fel$ is the element length, $E\fel$ stands for the Young's
modulus, possibly randomized in a binary sense, and $A\fel$ for the
cross-section area. The associated generalized element strain
$\eV\fel$ and stress $\sV\fel$ are defined as the bar's elongation
$\Delta \ell\fel$ and the axial force $S\fel$, respectively.

\begin{figure}[h]
\centering
 \includegraphics{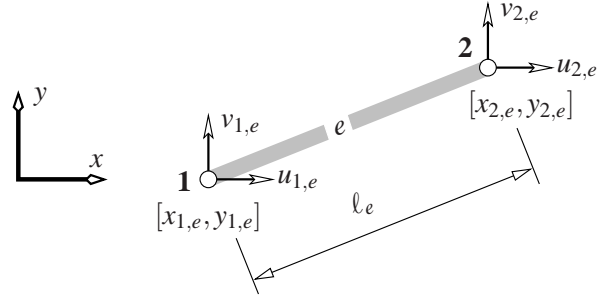}
\caption{Definition of the geometry and kinematics of a truss element.}
\label{fig:element_scheme}
\end{figure}

Assuming that $E\phs{1}\fel \geq E\phs{2}\fel$ for
$e=1,2,\ldots,\numel$, we introduce a dimensionless parameter in the form
\begin{eqnarray}
\relE_{\pma}( \PM\phs{11} ) 
= 
\frac{\sHF_{\pma}(\PM\phs{11}) 
- 
\HF\phs{2}}%
{\HF\phs{1} - \HF\phs{2}},
&&
0 \leq \relE \leq 1,
\end{eqnarray}
to aid the visualization of results. The
value of $\relE_{\pma}$ indicates the relative difference between the
estimated mean energy of a random system described by the $\PM\phs{11}$ statistics
and of deterministic structures, which solely consists of states
$r=1$ or $r=2$, yielding the deterministic energies
$\HF\phs{1}$ or $\HF\phs{2}$.

The regular lattice structure appearing in~\Fref{lattice_scheme} is
employed as a benchmark problem. The prescribed boundary conditions
include uniform tension and bending scenarios, imposed either using
nodal forces~(force control, $\FC$) or by prescribed nodal
displacements~(displacement control, $\DC$).

\begin{figure}[t]
\centerline{
 \includegraphics{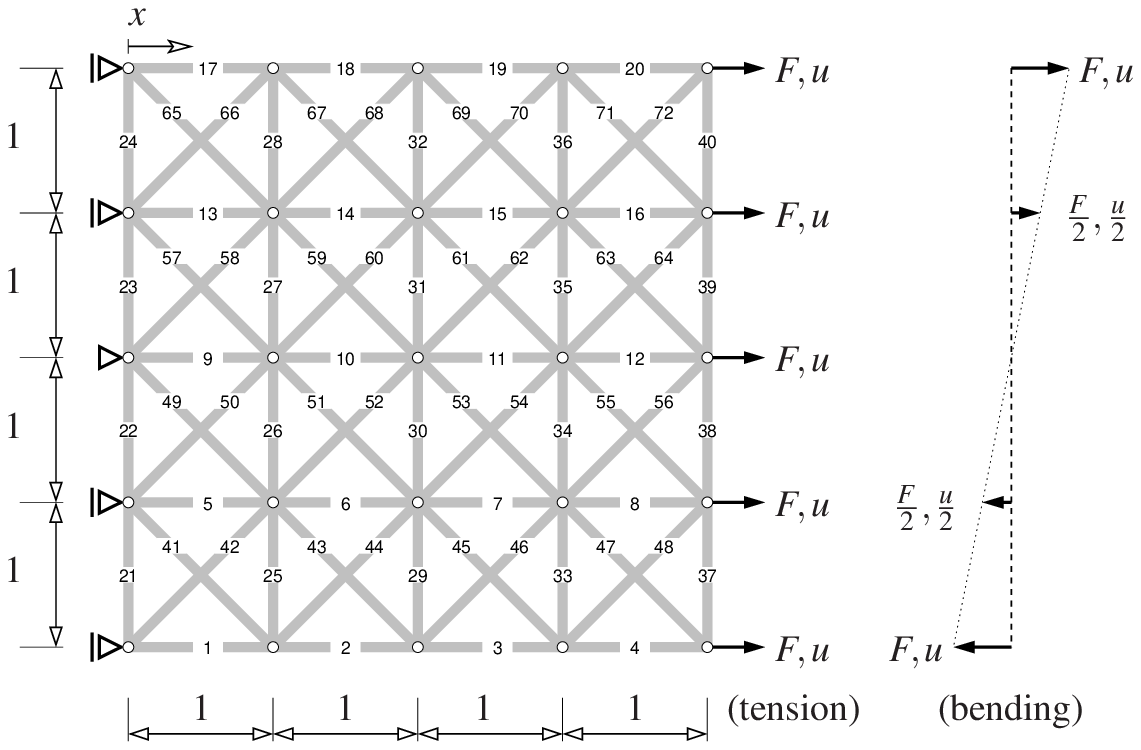}
}
\caption{Sketch of the lattice structure used as a benchmark problem.}
\label{fig:lattice_scheme}
\end{figure}

The results for second-order bounds presented hereafter are accompanied by
the statistically local counterparts of the
Voigt and Reuss type, determined for a deterministic problem with spatially
variable Young's moduli in the form
\begin{eqnarray}
E^{\V}\fel( \oppF\phs{1}\fel) 
& = &
\oppF\phs{1}\fel E\fel\phs{1} 
+ 
(1-\oppF\phs{1}\fel) E\fel\phs{2},
\\
E^{\R}\fel( \oppF\phs{1}\fel) 
& = &
\left( 
\frac{\oppF\phs{1}\fel}{E\fel\phs{1}}
+
\frac{1-\oppF\phs{1}\fel}{E\fel\phs{2}}
\right)^{-1}.
\end{eqnarray}
Finally, ensemble averages obtained by direct Monte-Carlo~($\MC$)
simulations with $N=10,000$ realizations are included as a reference.

\subsection{Structure with independent elements}\label{sec:example_rand_truss}
To demonstrate the performance of the non-local energetic bounds and estimates,
we start with the analysis of the simplest stochastic system with a closed-form
expression for the two-unit statistics. To this end, consider a binary structure
with the first phase assigned to each element independently with a probability
$\phi$. The associated second-order statistics are
\begin{equation}
\tppF\phs{11}_{ee'} 
=
\left\{
 \begin{array}{cl}
  \phi & \mbox{if } e=e', \\
  \phi^2 & \mbox{otherwise},
 \end{array}
\right.
\end{equation}
the system is therefore statistically uniform in the sense
of~\eqref{def_stat_uniform}. Notice that for the particular network shown
in~\Fref{lattice_scheme}, the cardinality of the ensemble space is $\card{\enspc} = 2^{72}
\doteq 4.72 \cdot 10^{21}$, whereas the computable two-unit
statistical characterization involves $\frac{72+1}{2}\cdot 72 = 2,628$
independent parameters only.

The contrast of phase stiffnesses is set to $E\phs{1} : E\phs{2} =
10:1$; the most restrictive upper Hashin-Shtrikman-Willis
bound~($\HSW_+$) therefore corresponds to a structure with
$E\phs{0}\fel = E\phs{1}$ and the optimal lower bound $\HSW_-$ is
obtained by setting $E\phs{0}\fel = E\phs{2}$. In addition, we include
variational estimates corresponding to the reference structure with
the generalized stiffness set to the Voigt upper
bound~($\HSW_{\sim\V}$) or the the Reuss bound~($\HSW_{\sim\R}$).

\begin{figure}
\centering
\subfigure[tension, $\FC$]{\includegraphics{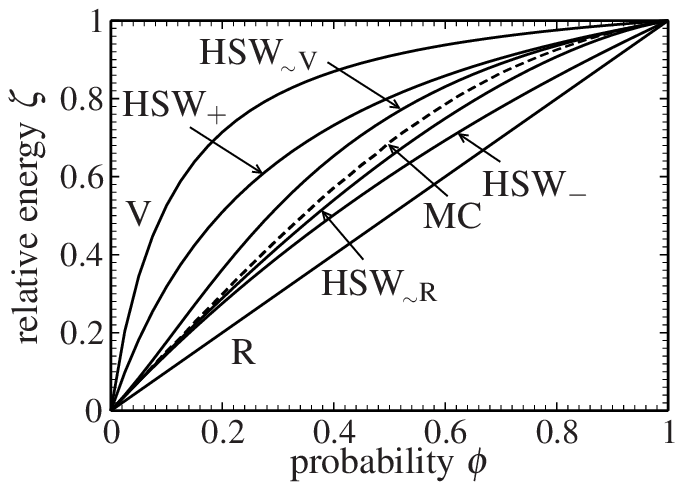}}
~
\subfigure[tension, $\DC$]{\includegraphics{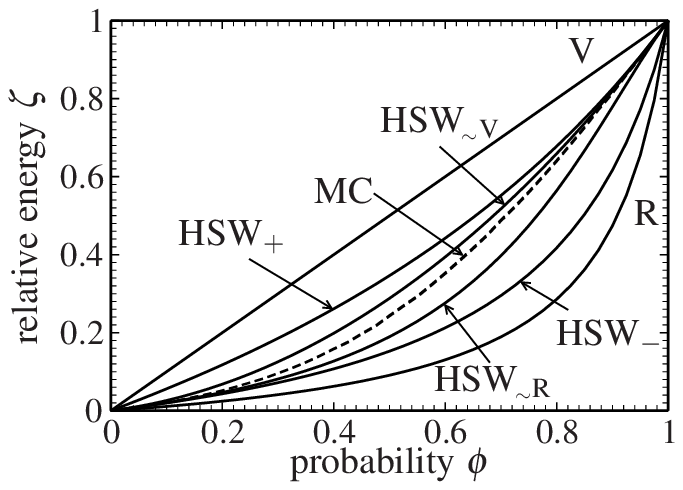}}
\\
\subfigure[bending, $\FC$]{\includegraphics{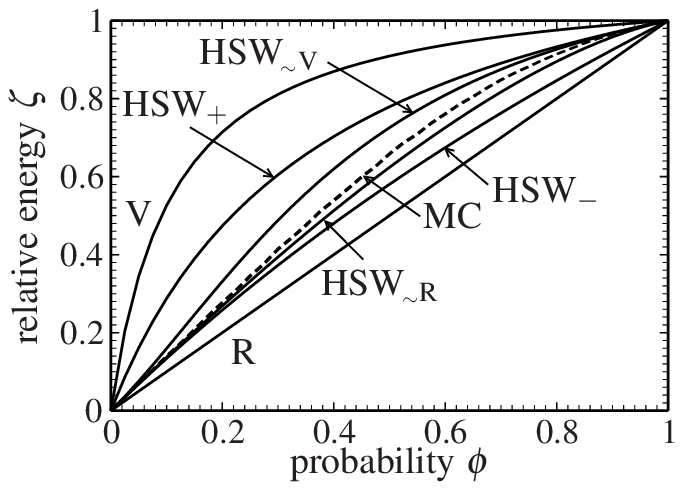}}
~
\subfigure[bending, $\DC$]{\includegraphics{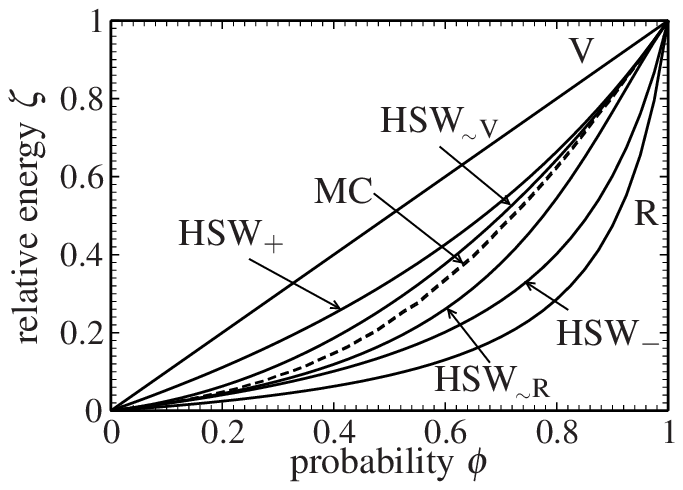}}
\caption{Energetics of the random truss structure with independent
  elements; (a)--(b)~tension, (c)--(d)~bending. The upper~($\HSW_+$)
  and lower~($\HSW_-$) Hashin-Shtrikman-Willis bounds are compared
  with the first-order Voigt~($\V$) and Reuss~($\R$) counterparts,
  variational estimates for the reference structure corresponding to
  the Voigt~($\HSW_{\sim\V}$) and the Reuss~($\HSW_{\sim\R}$) bound
  and direct $\MC$ simulations under displacement~($\DC$) and
  force~($\FC$) controlled loading.}
\label{fig:example1_energy}
\end{figure}

\Fref{example1_energy} presents the relative energies as a function of $\phi$ for all considered loading scenarios.
The~$\MC$ curves correspond to $99\%$ confidence intervals, thereby indicating
that a sufficient number of simulations was used to obtain reliable reference
data. Observe that, due to the adopted dimensionless representation, the shape
of the resulting curves is almost independent of the loading mode (tension or
bending). For all simulations, the $\HSW$ bounds substantially narrow the domain
defined by the first-order Voigt~($\V)$ and Reuss~($\R$) bounds while preserving
the concave/convex dependence of the mean potential
energy on the $\phi$ parameter for the displacement- or force-driven loads,
respectively. The increase in accuracy as a result of considering non-local
spatial statistics is especially pronounced for small and large values of
$\phi$, for which even the asymptotic behavior seems to be exactly reproduced by
the lower bound for the force-controlled conditions and the upper bound in the
kinematically constrained case. This observation is in an agreement with
analogous results for the Hashin-Shtrikman bounds for small values of the volume
fractions reported for isotropic elasticity~\citep{Roscoe:1973:ICE} or scalar
problems~\citep{Milton:1999:OGC}. Even better agreement between the~$\HSW$
predictions and $\MC$~simulations can be reached by employing the variational
estimates generated by setting the reference stiffness matrix equal to the
first-order bounds. Observe that for the particular system and loading
conditions considered here, the $\HSW_{\sim\R}$ and $\HSW_{\sim\V}$ estimates
effectively bound the reference data from above and below. As we are currently
unable to establish optimality of the $\HSW_+$ and $\HSW_-$
bounds, further generalization of this fact remains an open question.

The ``energetic'' displacements~\eqref{hs_mean_displ} as predicted by
the $\HSW$ bounds appear in~\Fref{example1_def_struct}. For both the
displacement and force driven cases, the average deformed shapes are
well reproduced by the bounds with, for example, the upper energetic
bound corresponding to smaller displacement values. This is fully
consistent with the fact that the upper bound represents the stiffest
response for a given load and two-unit statistics. Moreover, the
results comply with the intuitive trend suggesting a smaller
displacement range for a larger number of kinematic constraints.

\begin{figure}[h]
\begin{minipage}[c]{.475\textwidth}
\subfigure[$\DC$, maximum displacement scaled to $2$~m.]{\includegraphics*[height=45mm]{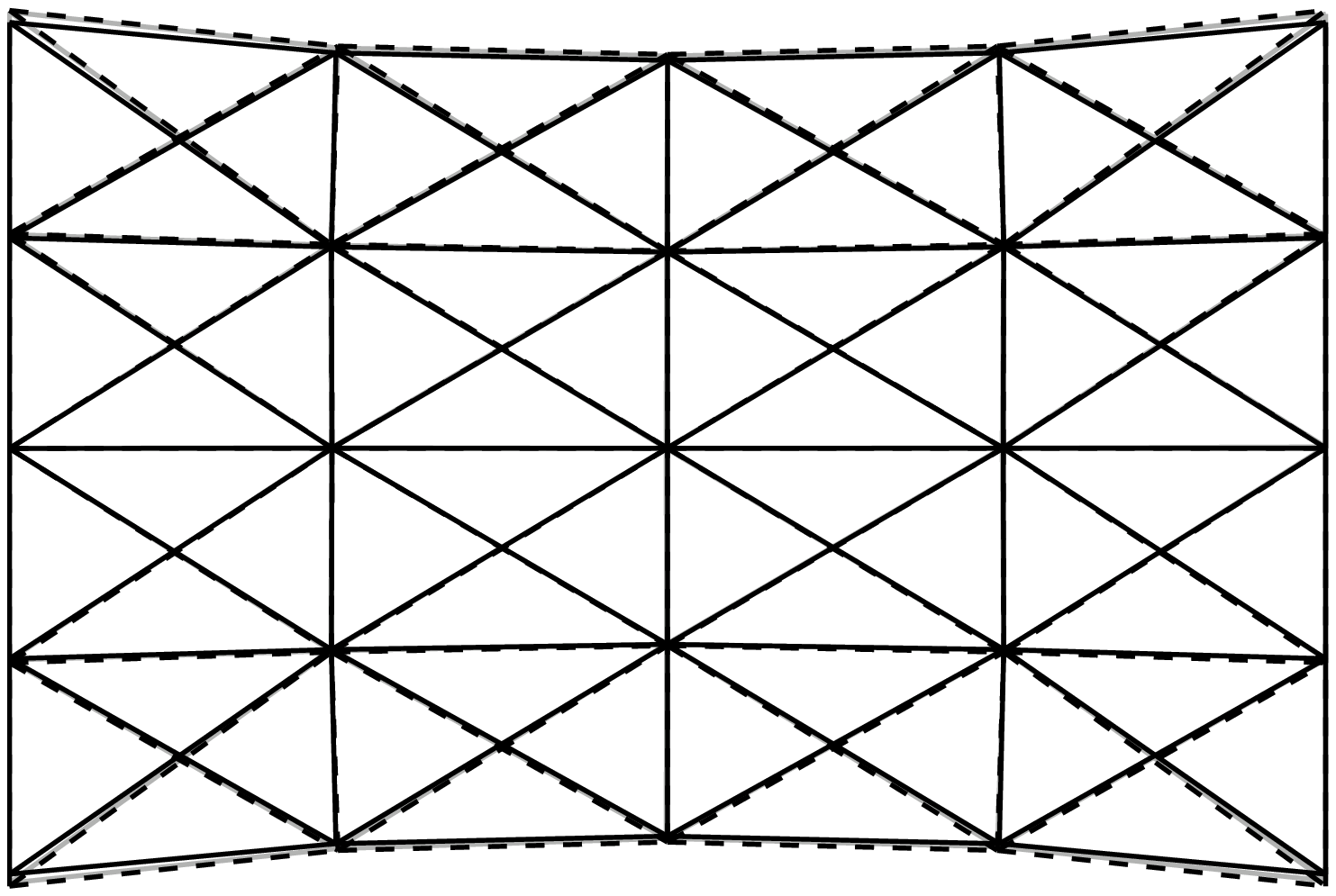}}
\end{minipage}
\hfill
\begin{minipage}[c]{.475\textwidth}
\subfigure[$\FC$, maximum displacement scaled to $0.25$~m.]{\includegraphics*[height=50mm]{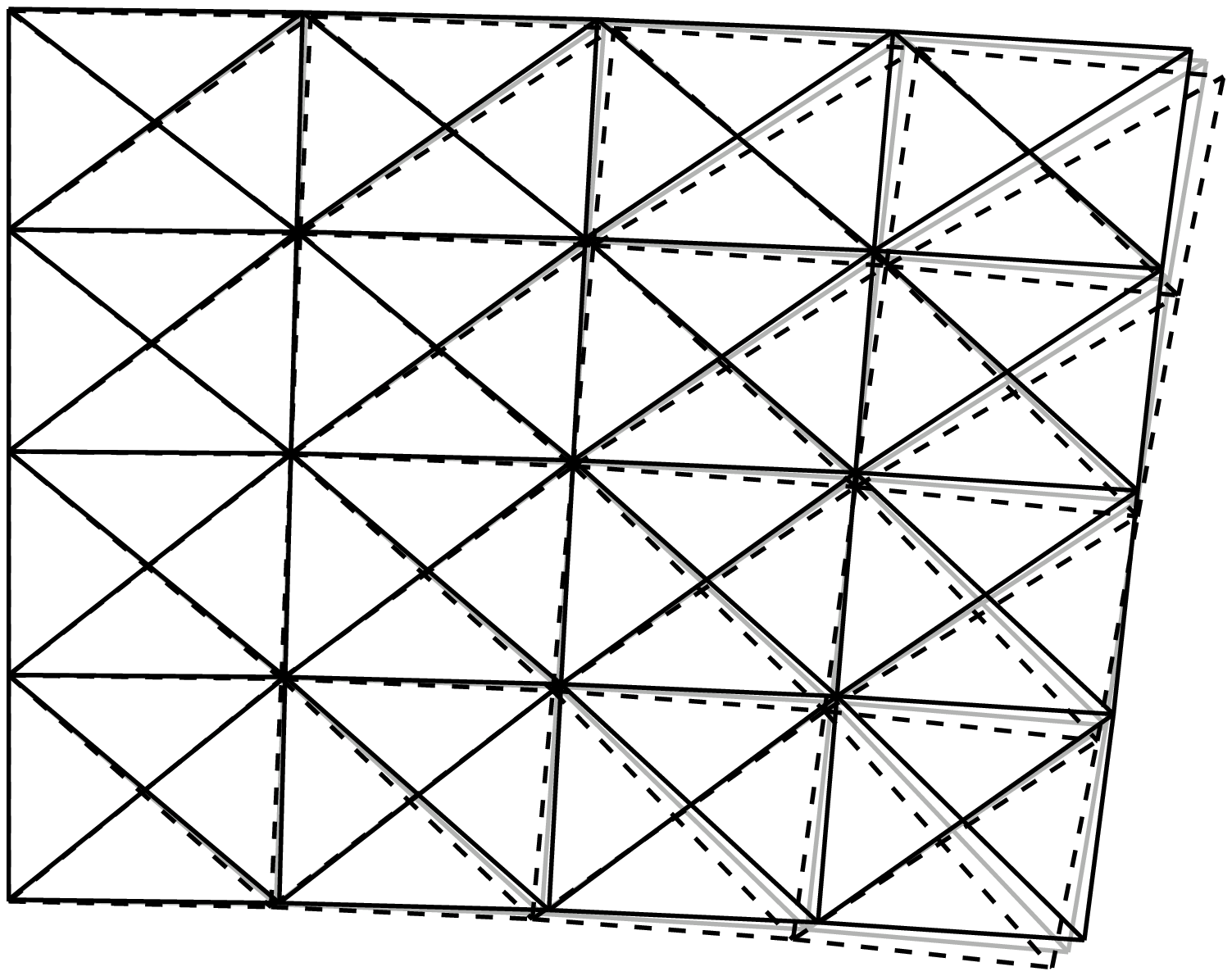}}
\end{minipage}
\caption{Deformed structures; (a)~tension, $\phi=0.3$,
  (b)~bending, $\phi = 0.6$ subject to displacement~($\DC)$ and
  force~($\FC$) driven loading; displacements corresponding to
  lower/upper bound are indicated by dashed/full lines and the results
  of Monte-Carlo simulations are depicted in gray.}
\label{fig:example1_def_struct}
\end{figure}

\subsection{Structure with spatially variable stiffness
distribution}\label{sec:example_non_uniform_truss}
Having demonstrated the accuracy and limitations of the bounds and
estimates for a system with independent units, we focus in the second example on
the effects of phase contrast and spatially variable distribution of material
properties. For simplicity, the statistical model is chosen to be identical to
the one introduced in the previous section; whereas the Young's moduli of
individual elements are now assumed in the form
\begin{eqnarray}
\frac{E\fel\phs{1}}{E\fel\phs{2}} 
=
1 
+
\left( 
\rho - 1
\right)
\frac{x_e}{4},
&&
e = 1, 2, \ldots, \numel,
\end{eqnarray}
where $x_e = (x_{1,e} + x_{2,e})/2$ denotes the $x$-coordinate of the $e$-th
element's center, recall~\Fref{lattice_scheme}, and $\rho \geq 1$ is the maximum
phase contrast.

\begin{figure}
\centering
\subfigure[$\FC, \rho=5$]{\includegraphics{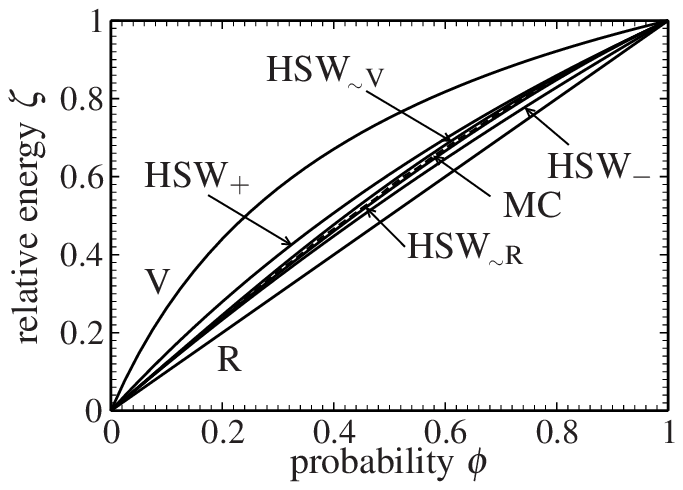}}
~
\subfigure[$\DC, \rho=5$]{\includegraphics{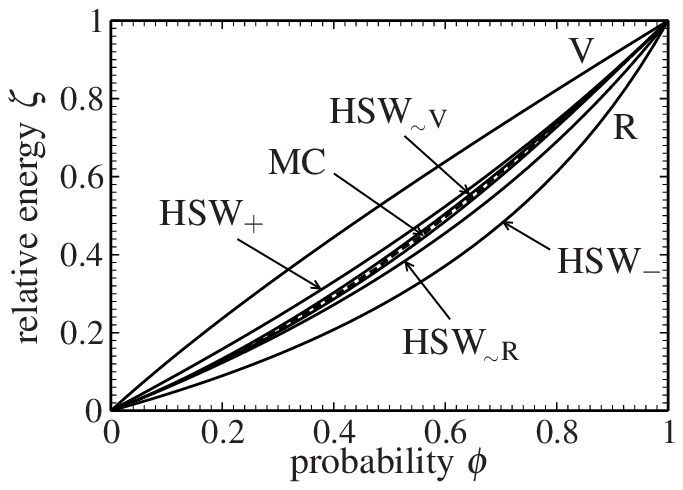}}
\\
\subfigure[$\FC, \rho=500$]{\includegraphics{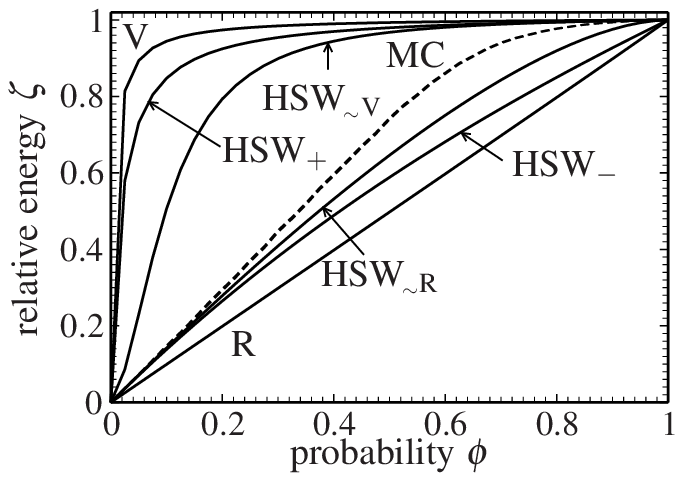}}
~
\subfigure[$\DC, \rho=500$]{\includegraphics{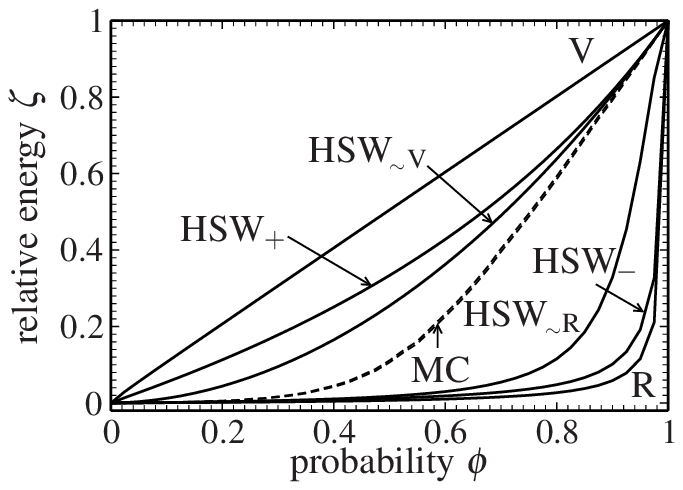}}
\caption{Energetics of the random truss structure with spatially
variable stiffness distribution subject to bending; (a)--(b)~phase
contrast $\rho=5$, (c)--(d)~phase contrast $\rho=500$. The upper~($\HSW_+$)
and lower~($\HSW_-$) Hashin-Shtrikman-Willis bounds are compared with the
first-order Voigt~($\V$) and Reuss~($\R$) counterparts,
  variational estimates for the reference structure corresponding to the
  Voigt~($\HSW_{\sim\V}$) and the Reuss~($\HSW_{\sim\R}$) bound and direct $\MC$
  simulations under displacement~($\DC$) and force~($\FC$) controlled loading.}
\label{fig:example3_energy}
\end{figure}

The resulting energetic bounds and estimates appear in~\Fref{example3_energy}
for the structure subject to bending for low~($\rho = 5$) and high~($\rho =
500$) phase contrasts, respectively. We observe that in the low-contrast case,
the region defined by the $\HSW_-$ and $\HSW_+$ bounds is substantially smaller
than in the previous example, whereas the $\HSW_{\sim\R}$ and $\HSW_{\sim\V}$
estimates almost coincide with the $\MC$ simulation data. For increasing values
of $\rho$, the accuracy of the bounds and estimates deteriorates; still the
statistically non-local quantities provide an improvement over the first-order
approaches. Moreover, analogously to \Sref{example_rand_truss}, they correctly
reproduce the asymptotic behavior for $\phi \rightarrow 0$ and $\phi \rightarrow
1$.

\subsection{Energetics of a weakest link fracture  model}\label{sec:weakest_link}

In the last example, we study the ability of the bounds and estimates
to capture the energetics of a simple randomized discrete fracture model with
two-unit probability matrices following from computer-generated
data.

In the deterministic setting,
following~\cite{Francfort:1993:SDE}, a damaged lattice is understood as a
two-state system with a ``healthy'' state~($\cF\fel\phs{1}=1, \cF\fel\phs{2}=0$)
and a fully-damaged state~($\cF\fel\phs{1}=0, \cF\fel\phs{2}=1$). As the fully
damaged elements cannot sustain any stress, we have $E\phs{2} : E\phs{1} = 0$.
The transition from state $1$ to state $2$ occurs when
\begin{equation}\label{eq:fract_cond}
\HF\fel
=
\half \dV\fel\trn \stM\fel \dV\fel
\geq 
\tR\fel,
\end{equation}
where $\stM\fel = \kM_{\el}^{\mathsf{T}} \mM\fel \kM\fel$ and
$\tR\fel$ denote the stiffness matrix and an energy threshold of the $\el$-th
element, respectively.

\begin{table}
\begin{enumerate}
  \item Input: probability distribution	of thresholds $\tR\fel$ for
  $e=1,2,\ldots, \numel$, loading scenario $f(t) \overline{\dV}$ with $f(0)=0,
  f(1)=1$ and $f$ being monotonic, number of realizations $N$ and sampling times
  $ 0 = \tau_0 < \tau_1 < \ldots < \tau_M = 1$ 
 \item For $i=1, 2, \ldots, N$
 \begin{enumerate}
   \item initialize $\vek{\phi} = \emptyset$, $\vek{\tau}=\emptyset$ and
   $\mtrx{\chi}^{[i]}(\tau_k) = \mtrx{1} \in \mathbb{R}^\numel$ for
   $k=1,2,\ldots M$

   \item set $E\fel = E$ and generate thresholds $\kappa\fel$ for
   $e=1,2,\ldots, \numel$

   \item Repeat
   \begin{enumerate}
     \item set $^{\vek{\phi}}\mtrx{\chi}^{[i]}( \tau_k ) = \mtrx{0}$ and $E\fel
     = 0$ for $e \in \mtrx{\phi}$ and $j \in \mtrx{\tau}$

     \item for current values of $E\fel$, compute element energies
     $\overline{H}_e$~\eqref{fract_cond} using displacements $\dV\fel$
     determined from~\eqref{equil_eqs} with $^{\fixed} \dV = \overline{\dV}$;
     $\el=1,2,\ldots,\numel$

     \item set $\vek{\varepsilon} = \{ e = 1, 2, \ldots, \numel :
     \overline{H}_e > 0 \}$
     \item compute $t_e = f^{-1}\left( \sqrt{\kappa\fel^{[i]} /
     \overline{H}_e} \right)$ for $e \in \vek{\varepsilon}$
     \item set $t = \min_{\el \in \vek{\varepsilon}} \{ t\fel \}$  
     \item set $\vek{\phi} = \left\{ e \in \vek{\varepsilon} : t_e = t \right\}$
     \item set $\vek{\tau} = \left\{ k = 1, 2, \ldots, M : \tau_k
     \geq t \right\}$
   \end{enumerate}
   until $t > 1$ or $\vek{\varepsilon} = \emptyset$

 \end{enumerate}

\item using samples $\{ \cV^{[i]}(\tau_k) {\cV^{[i]}}\trn( \tau_k ) \}_{i=1}^N$,
compute $\PM_\MC\phs{11}( \tau_k)$ for $k=1,2,\ldots, M$

\item using samples $\{ H^{[i]}( \tau_k )\}_{i=1}^N$, compute
$\sHF^\MC(\tau_k)$ for $k=1,2,\ldots, M$, where $\HF^{[i]}(\tau_k) = \sum_{e=1}^\numel
   \HF\fel$, with $\HF\fel$ provided by~\eqref{fract_cond} using displacements
   $\dV\fel$ determined from~\eqref{equil_eqs} with $^{\fixed}\dV = f(\tau_k)
   \overline{\dV}$ and $E\fel = \cF^{[i]}\fel( \tau_k ) E$

\end{enumerate}
\caption{Conceptual implementation of the weakest link algorithm.}
\label{tab:weakest_link}
\end{table}

The stochastic damage evolution is simulated using a simple combination
of an ``event-by-event'' strategy and a direct Monte-Carlo algorithm as
outlined in Table~\ref{tab:weakest_link}. For every realization
of energy thresholds, a deterministic time-stepping procedure is executed. First,
the structure is subjected to the full displacements and the energies
corresponding to individual elements are extracted~(step ii.). Then, in steps
iii.--vii., the element(s) with the highest damage affinity is~(are) determined
and removed from the structure~(step i.). This is accompanied by the bookkeeping
of element deactivation times in terms of the auxiliary matrices $\cV^{[i]}(
\tau_k ), k=1,2,\ldots, M$. The procedure is repeated until the imposed loading sequence
is completed or until the structure loses its integrity; cf. the termination
conditions of step~(c). After the sampling phase is complete, quantities of
interest related to the damage statistics follow from post-processing of
the simulation results~(steps 3. and 4.); see also
e.g.~\citep{Sharif-Khodaei:2008:MBM,Alava:2006:SMF,Vellinga:2008:IBC} and
references therein for more details.

In particular, we consider the lattice shown in~\Fref{lattice_scheme}
subject to an imposed displacement, which is parametrized as
\begin{equation}
u(t) = u_{\max} \sqrt{t}, 
\quad 
0 \leq t \leq 1,
\end{equation}
leading to a linear scaling of the element energies~\eqref{fract_cond}
with respect to the pseudo-time~$t$.

For simplicity, identical values of the Young's moduli, $E\phs{1}\fel
= E$, are assumed in the healthy state for every element and the
thresholds are taken as uniformly distributed independent random
variables ranging from $\tR\fel^-=\frac{1}{4} E\fel A\fel (\Delta
\ell\fel\crt)^2/\ell\fel$ to $\tR\fel^+=3 \tR\fel^-$ with the critical
elongation $\Delta \ell\fel\crt = 10^{-3} \ell\fel$. The target
displacement is set to $u_{\max} = 10^{-2}$ for uniform tension,
whereas for the bending case $u_{\max}=1.5 \cdot 10^{-2}$. The time
evolution of the system was sampled at $M=100$ uniformly distributed time
instants using $N=10,000$ realizations.

\begin{figure}[p]
\centering \includegraphics[scale=.925]{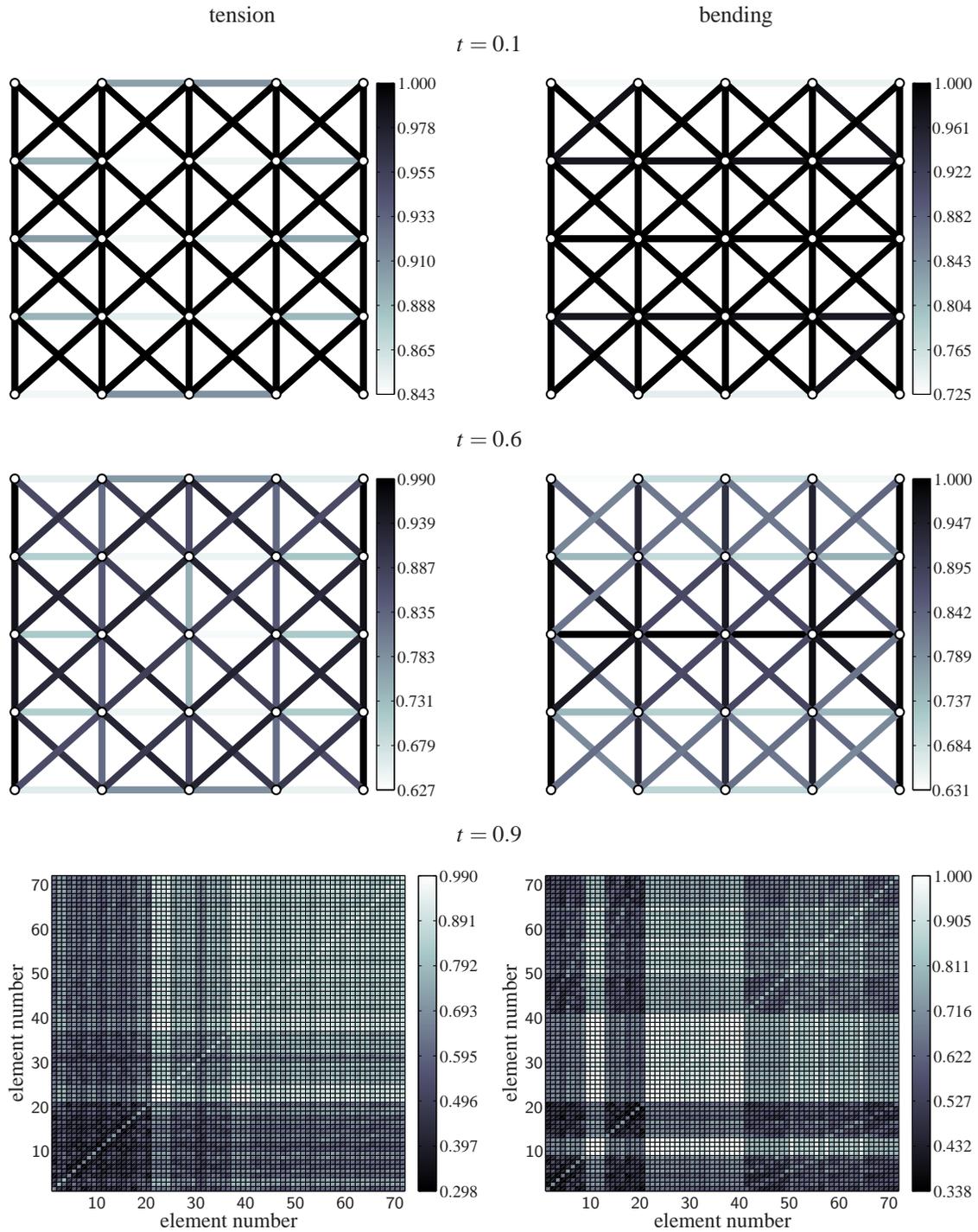}
\caption{Evolution of the one- and two-unit probability matrix
  $\PM_\MC\phs{11}$ entries as determined by Monte-Carlo
  simulations. The snapshots for $t=0.1$ and $t=0.6$ show one-unit
  probabilities, whereas complete two-unit probability matrix is
  plotted for $t = 0.9$.}
\label{fig:example2_damage_evolution}
\end{figure}

Snapshots of the damage progress quantified in terms of one- and two-unit
probabilities appear in~\Fref{example2_damage_evolution} for both the uniform
tension and bending scenarios. Note that the diagonal entries of the
$\PM\phs{11}( t )$ matrix provide, in the current setting, the survival
probability for each element at time~$t$. The off-diagonal values correspond to
the simultaneous survival probabilities for two elements. 

In the case of uniform tension, the damage distribution appears to be rather
diffuse during the whole loading procedure. Low survival probabilities are
localized in the horizontal links in the corners of the network and in the
structure's interior, as these exhibit the highest failure affinity for the
deterministic solution with $2 \tR\fel = \tR\fel^- + \tR\fel^+$~(pseudo-time $t
= 0.1)$.\footnote{Therefore, in the terminology introduced
by~\Eref{def_stat_uniform}, the initially statistically uniform structure evolves
to a statistically non-uniform system.} The slight non-symmetry of the one-unit
probabilities visible in~\Fref{example2_damage_evolution} results from the
finite number of realizations used to sample the statistics. With increasing
load, the remaining horizontal elements start to fail due to the energy
redistribution after failure of the first group of weakest links, leading to a
decrease of the survival probabilities for the interior elements~($t=0.6$). In
the case of bending, on the other hand, the damage is triggered mainly in the
upper and lower rows of elements and subsequently propagates to internal links.
Additional details on the simultaneous survival probabilities are given by the
full two-unit probability matrices, shown in~\Fref{example2_damage_evolution}
for $t=0.9$. Recalling element numbering introduced in~\Fref{lattice_scheme},
the plots reveal, among other things, simultaneous survival probabilities close
to one for vertical element groups $21$--$24$ and $37$--$40$, which connects the
nodes where the load is imposed, and horizontal elements $9$--$10$, at the
neutral axis, in the case of bending. It follows from these results that the
second-order statistics correctly capture the dominant fracture mechanisms,
thereby providing a well-founded statistically non-local damage
parameter for stochastic damage theories.

The evolution of the normalized stored energy $\relE$ is depicted
in~\Fref{example2_energy_evolution}. The choice of reference
structures to generate the bounds or estimates is identical to that in
the previous example. Note that the lower bounds and the
$\HSW_{\sim\R}$ estimate are identically equal to zero due to the zero
stiffness assigned to the weaker state. The overall character of the
energy evolution is consistent with the previous discussion. In
particular, two failure modes are clearly visible for the bending
problem, whereas the gradual decrease in energy for uniform tension
corresponds to more a diffuse damage character. Similarly
to~\Sref{example_rand_truss}, the first-order Voigt bounds correctly
predict the trend of the energy reduction, but under predicts its
magnitude. The $\HSW$ bound remains highly accurate until $t \approx
0.2$. A certain discrepancy, however, is observed for the limit value
of the dimensionless energy, e.g. $\relE^{\HSW_+} \doteq 0.24$ instead
of the reference $\relE^{\MC} \approx 0$ for uniform tension,
cf.~\Fref{example2_energy_evolution}(a). This difference again arises
due to the fact that the upper $\HSW$ bound corresponds, in the
ensemble average sense, to the response of the stiffest structure
compatible with the two-unit statistics shown
in~\Fref{example2_damage_evolution}, for which the minimum probability
reaches $\approx 30\%$ for both loading sequences. The results of the
$\MC$~simulations, on the other hand, represent one particular
stochastic system implicitly defined by the randomization procedure
and the event-by-event solution algorithm. Nevertheless, the added
value of the second-order relations becomes apparent when considering
the Voigt prediction $\relE^{\V} \doteq 0.74$. Slightly more accurate
values of the residual relative energy can be generated using the
$\HSW_{\sim\V}$ estimate, for which we
obtain~$\relE^{\HSW_{\sim\V}}\doteq 0.21$. The general character of
these conclusions is further supported by the analogous results of the
bending mode shown in~\Fref{example2_energy_evolution}(b).

\begin{figure}[h]
\subfigure[]{\includegraphics{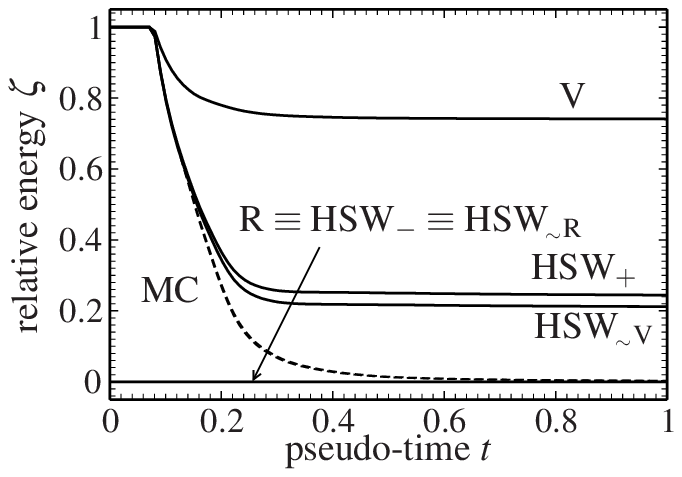}}
\hfill
\subfigure[]{\includegraphics{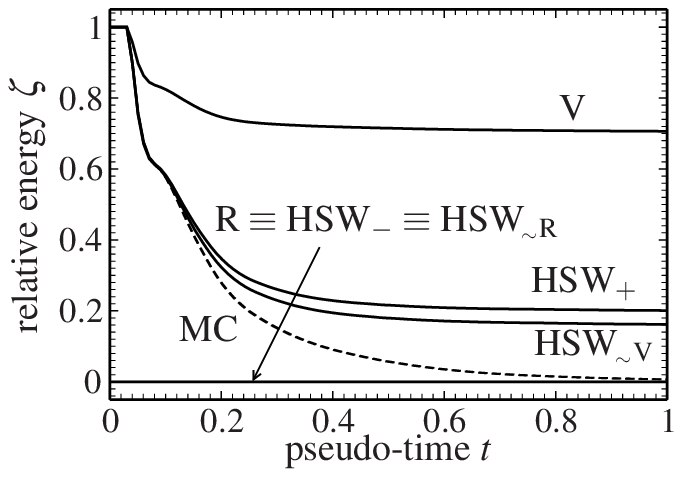}}
\caption{Energetics of the weakest link model; (a)~tension,
  (b)~bending.}
\label{fig:example2_energy_evolution}
\end{figure}

\section{Conclusions}\label{sec:concl}
In this work, variational bounds and estimates of the
Hashin-Shtrikman-Willis~($\HSW$) type for two-phase random structures
have been derived and verified against the results of direct
Monte-Carlo~($\MC$) simulations. The most important findings can be
summarized as follows:
\begin{itemize}

\item When applied to discrete structures instead of continua, the
  derivation of the $\HSW$ principles becomes reasonably
  straightforward and requires only elements of matrix structural
  analysis and linear algebra.

\item The variational framework naturally incorporates general
  statistically non-uniform systems, including the specification of
  the set of all admissible two-unit probability matrices. Thus, the
  developed statistically non-local theory is able to treat the
  statistics of systems with independent and highly correlated units in a unified
  manner.

\item The second-order bounds and estimates provide a computationally
  feasible alternative to direct $\MC$ simulations.

\item When applied to the displacement-driven damage problems, the
  upper bound on the stored energy delivers a reasonably accurate approximation
  to the total potential energy at the damage initiation
  and propagation without any adjustable parameters. When complemented with an
  accurate representation of the irreversible energy dissipation during the
  damage process, the scheme has the potential to provide a consistent
  variational model for damage evolution in discrete media.

\end{itemize}

The developed statistically non-local discrete theory
can also provide a convenient starting point for the coarsening towards an equivalent
continuum representation. The already announced
applications, i.e. a variational approach to
deterministic and stochastic damage mechanics of finite-sized lattices
and incorporation of higher-order statistics, will explored and reported separately in future publications.

\appendix

\section{Derivation of condensed energy}\label{sec:cond_energy}
%
Recall that the expression for the condensed energy follows from the
value of the Hashin-Shtrikman function~\eqref{HS_energy_func_def} with
the displacement vector optimally adjusted to a trial generalized
polarization stress~$\trial{\tV}$. In particular, evaluating the
energy for the displacement~${}^{\free}\sopt{\dV}$ determined
from~\eqref{HS_displ_split} yields, in a partitioned format:
\begin{eqnarray*}
\UF( {}^{\free}\sopt{\dV}, \trial{\tV}; \rl ) 
& = &
\half
\myMatrix{c}{%
 {}^{\free}\dV\cmp{0}
 + 
 {}^{\free}\sopt{\dV}\cmp{1}
 \\
 {}^{\fixed}\dV
}^{\sf T}
\myMatrix{cc}{%
 {}^{\free\free}\stM\phs{0} & {}^{\free\fixed}\stM\phs{0} \\
 {}^{\fixed\free}\stM\phs{0} & {}^{\fixed\fixed}\stM\phs{0}
}
\myMatrix{c}{%
 {}^{\free}\dV\cmp{0}
 + 
 {}^{\free}\sopt{\dV}\cmp{1}
 \\
 {}^{\fixed}\dV
}
\\
& - &
\left( 
 {}^{\free}\dV\cmp{0}
 + 
 {}^{\free}\sopt{\dV}\cmp{1}
 \right)\trn
~{}^{\free}\fV
+
\trial{\tV}\trn
\myMatrix{cc}{%
{}^{:\free}\kM 
& 
{}^{:\fixed}\kM 
}
\myMatrix{c}{%
 {}^{\free}\dV\cmp{0}
 + 
 {}^{\free}\sopt{\dV}\cmp{1}
 \\
 {}^{\fixed}\dV
} 
\\
& + & 
\half
\trial{\tV}
\left( 
 \mM\phs{0} - \mM(\rl)
\right)\inv
\trial{\tV}.
\end{eqnarray*}

A suitable re-arrangement of terms appearing in the previous relation
together with the optimality conditions~\eqref{system_1}
and~\eqref{system_2} leads to
\begin{eqnarray*}
\UF( {}^{\free}\sopt{\dV}, \trial{\tV}; \rl ) 
& = & 
\overbrace{%
\half
\myMatrix{c}{%
 {}^{\free}\dV\cmp{0}
 \\
 {}^{\fixed}\dV
}^{\sf T}
\myMatrix{cc}{%
 {}^{\free\free}\stM\phs{0} & {}^{\free\fixed}\stM\phs{0} \\
 {}^{\fixed\free}\stM\phs{0} & {}^{\fixed\fixed}\stM\phs{0}
}
\myMatrix{c}{%
 {}^{\free}\dV\cmp{0}
 \\
 {}^{\fixed}\dV
}
-
{}^{\free}\dV\cmp{0}
\trn 
{}^{\free}\fV
}^{=\HF\phs{0} \text{~by definition}}
\\
& + & 
\trial{\tV}\trn
\overbrace{%
\myMatrix{cc}{%
{}^{:\free}\kM 
& 
{}^{:\fixed}\kM 
}
\myMatrix{c}{%
 {}^{\free}\dV\cmp{0}
 \\
 {}^{\fixed}\dV
}
}^{=\eV\cmp{0} \text{~by definition}}
+
\half
\trial{\tV}
\left( 
 \mM\phs{0} - \mM(\rl)
\right)\inv
\trial{\tV}
\\
& + & 
\half
\trial{\tV}\trn
\overbrace{%
{}^{:\free}\kM 
\,{}^{\free}\sopt{\dV}\cmp{1}
}^{=-\gM\phs{0}\trial{\tV} \text{~by~\Eref{gamma_0_def}}}
+
\half
{}^{\free}\sopt{\dV}\cmp{1}
\trn
\overbrace{
\left(
 \stM\phs{0} 
 {}^{\free}\sopt{\dV}\cmp{1}
 + 
 {}^{:\free}\kM\trn
 \trial{\tV}
\right)
}^{=\vek{0} \text{~by \Eref{system_2}}}
\\
& + & 
{}^{\free}\sopt{\dV}\cmp{1}
\trn
\overbrace{%
\left(
 \myMatrix{cc}{%
 {}^{\free\free}\stM\phs{0} & 
 {}^{\free\fixed} \stM\phs{0}
}
\myMatrix{c}{%
 {}^{\free}\dV\cmp{0}
 \\
 {}^{\fixed}\dV
}
-
{}^{\free}\fV
\right)}^{=\vek{0} \text{~by \Eref{system_1}}}
\end{eqnarray*}
which coincides with the expression for the condensed energy as
appearing in~\Eref{cond_energy}. Note that similar development for the
continuous case is available in, e.g.,~\cite{Willis:1977:BSC} and
in~\cite{Willis:1981:VRM}, with the FE-discretization treated
in~\cite{Luciano:2005:FE}.

\section*{Acknowledgments}
We would like to thank an anonymous reviewer for numerous valuable
remarks and suggestions to improve the clarity of the paper. The work of JZ was
supported by the Marie-Curie fellowship, project No.~MEIF-CT-2005-024392, and by
research projects GA~106/08/1379~(Czech Science Foundation) and MSM
684077003~(Ministry of Education, Youth and Sports of the Czech Republic).

\end{document}